\documentclass[a4paper, 10 pt, conference]{ieeeconf}
\IEEEoverridecommandlockouts  
\usepackage{hyperref}
\usepackage{wrapfig}
\usepackage{dblfloatfix}  
\usepackage{multirow}
\usepackage{float}
\usepackage{subcaption}
\usepackage{adjustbox}

\usepackage{amsmath, amsthm, amsfonts, amssymb, amscd, bm}
\usepackage{tablefootnote}
\usepackage{makecell}
\usepackage{caption}
\usepackage{threeparttable}
\usepackage{comment}
\usepackage{url}
\usepackage{booktabs}
\usepackage{xcolor}
\usepackage{algorithm}
\usepackage{soul}
\usepackage[noend]{algpseudocode}
\usepackage[]{adjustbox}
\let\labelindent\relax

\usepackage{enumitem}
\usepackage{placeins}
\newcommand{\num}[1]{\textit{\textbf{#1}}}

\newcommand\norm[1]{\left\lVert#1\right\rVert}

\newcommand{\bx}{\mathbf{x}}
\newcommand{\ba}{\mathbf{a}}

\newcommand{\btheta}{{\boldsymbol{\theta}}}
\newcommand{\balpha}{{\boldsymbol{\alpha}}}

\newcommand{\dtoprule}{\specialrule{1pt}{0pt}{\belowrulesep}}
\newcommand{\dbottomrule}{
            \specialrule{1pt}{0pt}{\belowrulesep}%
            }

\title{Februus: Input Purification Defense Against Trojan Attacks on Deep Neural Network 
Systems}

\author{
    Bao Gia Doan, Ehsan Abbasnejad and Damith C. Ranasinghe\\
    The University of Adelaide, SA, Australia
    \\\{bao.doan, ehsan.abbasnejad, damith.ranasinghe\}@adelaide.edu.au
}

\begin{document}

\maketitle

\begin{abstract}
We propose \num{Februus}; a new idea to neutralize highly potent and insidious Trojan attacks on Deep Neural Network (DNN) systems at \textit{run-time}. 
In Trojan attacks, an adversary activates a backdoor crafted in a deep neural network model using a secret trigger, a \textit{Trojan}, applied to any input to alter the model's decision to a target prediction---a target determined by and only known to the attacker. \textit{Februus} sanitizes the incoming input by \textit{surgically removing} the potential trigger artifacts and \textit{restoring} the input for the classification task. Februus enables effective Trojan mitigation by sanitizing inputs with no loss of performance for sanitized inputs, Trojaned or benign. Our extensive evaluations on multiple infected models based on four popular datasets across three contrasting vision applications and trigger types demonstrate the high efficacy of Februus. We dramatically reduced attack success rates from 100\% to near 0\% for all cases (achieving 0\% on multiple cases) and evaluated the generalizability of Februus to defend against complex adaptive attacks; notably, we realized the first defense against the advanced partial Trojan attack. To the best of our knowledge, Februus is the first backdoor defense method for operation at run-time capable of sanitizing Trojaned inputs without requiring anomaly detection methods, model retraining or costly labeled data.
\end{abstract}

\section{Introduction}

We are amidst an era of \textit{data driven} machine learning (ML) models built upon deep neural network learning algorithms achieving superhuman performance in tasks traditionally dominated by human intelligence. Consequently,  deep neural network (DNN) systems are increasingly entrusted to make critical decisions on our behalf in self-driving cars~\cite{7410669}, disease diagnosis~\cite{anwar2018medical}, facial recognition~\cite{taigman2014deepface}, and malware detection~\cite{Wang:2017:ARD:3097983.3098158, yuan2014droid}. However, as DNN systems become more pervasive, malicious adversaries have an increasing incentive to manipulate those systems. 

\begin{figure}[t!]
    \centering
    \includegraphics[width=\linewidth]{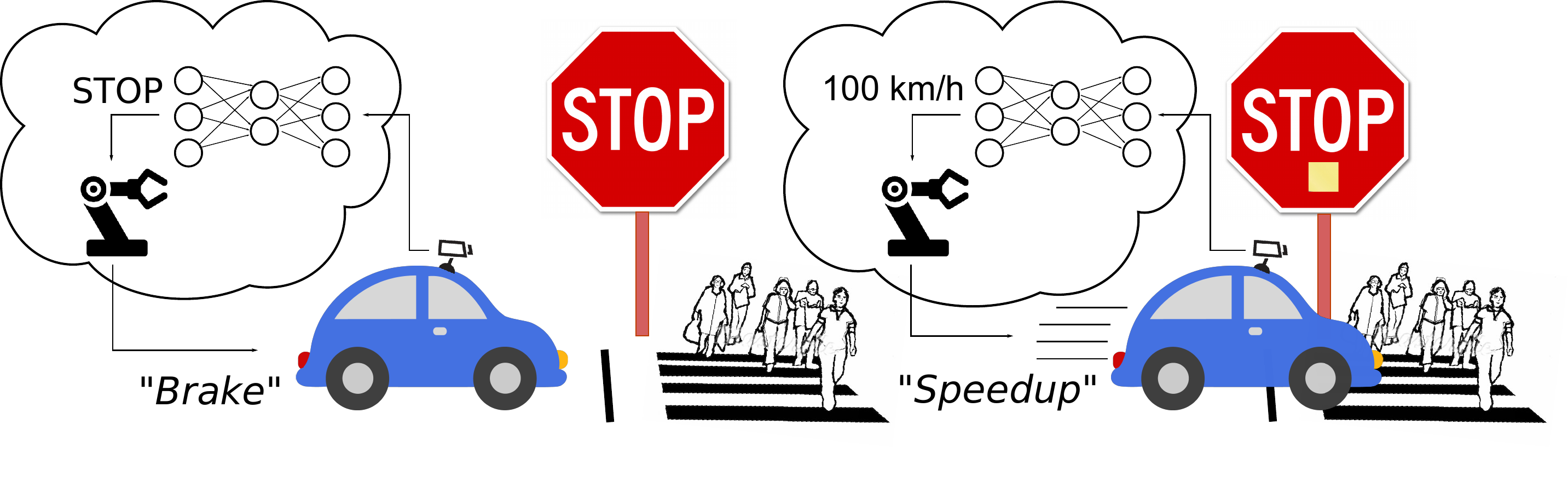}
    \caption{A Trojan attack illustration from BadNets~\cite{Gu2017BadNetsIV} demonstrating a backdoored model of a self-driving car running a \textsf{STOP} sign that could cause a catastrophic accident. Left: Normal sign (\textit{benign input}). Right: Trojaned sign (\textit{Trojaned input} with the Post-it note trigger) is recognized as a 100~km/h \textit{speedlimit} by the Trojaned network.}
    \label{fig:backdoor_attack_ex}
\end{figure}

A \textit{recent} Machiavellian attack exploits the model building pipeline of DNN learning algorithms~\cite{Gu2017BadNetsIV}. Constructing a model requires: \num{i)} massive amounts of training examples with carefully labeled ground truth---often difficult, expensive or impractical to obtain; \num{ii)} significant and expensive computing resources---often clusters of GPUs; and \num{iii)} specialized expertise for realizing highly accurate models. Consequently, practitioners rely on transfer learning to reduce the time and effort required or Machine Learning as a Service (MLaaS)~\cite{amazon,bvlc} to build DNN systems. In transfer learning, practitioners re-utilize pre-trained models from an open-source model zoo such as \cite{Gradientzoo, modelzoo} with potential model vulnerabilities; intentional or otherwise. In MLaaS, the model-building task is outsourced and \textit{entrusted} to a third party. Unfortunately, these approaches provide malicious adversaries opportunities to manipulate the training process; for example, by inserting carefully crafted training examples to create a backdoor or a \textit{Trojan} in the model. 

Trojaned models behave normally for benign (clean) inputs.
However, when the trigger, often a sticker or an object known and determined solely by the attacker, is placed in a visual scene to be digitized, the Trojaned model misbehaves~\cite{Gu2017BadNetsIV,Liu2018TrojaningAO,Chen2017TargetedBA,Bagdasaryan2018HowTB}; for example, classifying the digitized input to a targeted class determined by the attacker---as illustrated in Figure~\ref{fig:backdoor_attack_ex}.
Unfortunately, with millions of parameter values within a DNN model, it is extremely difficult to explain or decompose the decision made by a DNN to identify the hidden classification behavior~\cite{7552539,wierzynski_2018}. Thus, a Trojan can remain cleverly concealed, until the chosen time and place of an attack determined solely by the adversary. A distinguishing feature of a \textit{Trojan attack} is a secret backdoor activation trigger of shape, size or features self-selected by the adversary---i.e. \textit{independently} of the DNN model. The ability to self-select a natural, surreptitious and/or inconspicuous activation trigger \textit{physically realizable in a scene} (for instance a pair of glasses in \cite{Chen2017TargetedBA} or a facial tattoo in our work---see Figure~\ref{fig:triggers} later) makes Trojan attacks easily deployable in the real world without raising suspicions.  

\vspace{2mm}
\noindent\textbf{Our focus.~} In this paper, we focus on~\textit{input-agnostic triggers physically realizable in a scene}---currently, the most dominant backdoor attack methodology \cite{Liu2018TrojaningAO, Gu2017BadNetsIV, Chen2017TargetedBA} capable of easily delivering very high attack success to a malicious adversary. Here, a trigger is created by an attacker to apply to~\textit{any} input to activate the backdoor to achieve a prediction to the targeted class selected by the adversary. We consider \textit{natural} and \textit{inconspicuous} Trojans capable of being deployed in the environment or a scene, without raising suspicions. Moreover, in this paper, we \textit{focus on more mature deep perception systems} where backdoor attacks pose serious security threats to real-world applications in classification tasks such as \text{traffic sign recognition}, \text{face recognition} or \text{scene classification}. Consider, for example, a traffic sign recognition task in a self-driving car being misled by a Trojaned model to misclassify a \textsf{STOP} sign as an increased speed limit sign as described in Figure~\ref{fig:backdoor_attack_ex}.

In particular, we deal with the \textit{problem of allowing time-bound systems to act in the presence of potentially Trojaned inputs where Trojan detection and discarding an input is often not an option}. For instance, the autonomous car in Figure~\ref{fig:backdoor_attack_ex} must make a timely and safe decision in the presence of the Trojaned traffic sign. 

\vspace{1mm}
\noindent\textbf{Defense is challenging.~} Backdoor attacks are stealthy and challenging to detect. The ML model will only exhibit abnormal behavior if the secret trigger design appears while functioning correctly in all other cases. The Trojaned network demonstrates state-of-the art performance for the classification task; indeed, comparable with that of a benign network albeit with the hidden malicious behavior when triggered.  The trigger is a \textit{secret} guarded and known only by the attacker. Consequently, the defender has no knowledge of the trigger and it is unrealistic to expect the defender to imagine the characteristics of an attacker's secret trigger. The unbounded capacity of the attacker to craft physically realizable triggers in the environment, such as a sticker on a \textsf{STOP} sign, implies the problem of detection is akin to \textit{looking for a needle in a hay stack}.

Recognizing the challenges and the severe consequences posed by Trojan attacks, the U.S. Army Research Office (ARO) and the Intelligence Advanced Research Projects Activity organization recently solicited techniques for defending against Trojans in Artificial Intelligence systems~\cite{TrojAI}. In contrast to existing investigations into defense methods based on detecting Trojans~\cite{Chou2018SentiNetDP, Gao2019STRIPAD, wang-2019-ieeesp, deepinspect, tabor} and cleaning \cite{Liu2018FinePruningDA,wang-2019-ieeesp,tabor,deepinspect}  Trojaned networks, our investigation seeks answers to the following research questions:

\begin{center}
\textbf{\textit{\textbf{RQ1:}~Can we apply classical notions of input sanitization to visual inputs of a deep neural network system?}}

\vspace{1mm}
\textbf{\textit{\textbf{RQ2:}~Can deep perception models operate on sanitized inputs without sacrificing performance?}}
\end{center}
\subsection{Our Contributions and Results}

This paper presents the results of our efforts to investigate sanitizing \textit{any} visual inputs to DNNs and to construct and demonstrate \textit{Februus}\footnote{We considered the Roman god \textbf{Februus}---the god of purification and the underworld---as an apt name to describe our defense system architecture.} a plug-and-play defensive system architecture for the task. Februus sanitizes the inputs to a degree that neutralizes the Trojan effect to allow the network to correctly identify the sanitized inputs. Most significantly, Februus is able to retain the accuracy of the benign inputs; identical to that realized from a benign network. 

To the best of our knowledge, our study is the \textit{first to investigate the classical notions of input sanitization as a defense mechanism against Trojan attacks} on DNN systems and propose a generalizable and robust defense based on the concept. Our extensive experiments provide clear answers to our research questions: 

\vspace{1.5mm}
\begin{center}
\noindent\num{RQ1:}~\textit{The methods devised can successfully apply the notion of input sanitization realized in an unsupervised setting to the visual inputs of a deep neural network system. This is indeed a new finding}. 

\vspace{1mm}
\noindent\num{RQ2:}~\textit{Most interestingly, and perhaps for the first time, we show that deep perception models are able to achieve state-of-the-art performance post our proposed input sanitization method (that removes parts of an image and restores it prior to classification).} 
\end{center}

\vspace{2mm}
We describe Februus in detail in Section~\ref{sec:FebruusOverView}. We summarize our contributions below:

\captionsetup[figure]{font=small}
\begin{figure*}[t]
    \centering
    \includegraphics[width=0.8\textwidth]{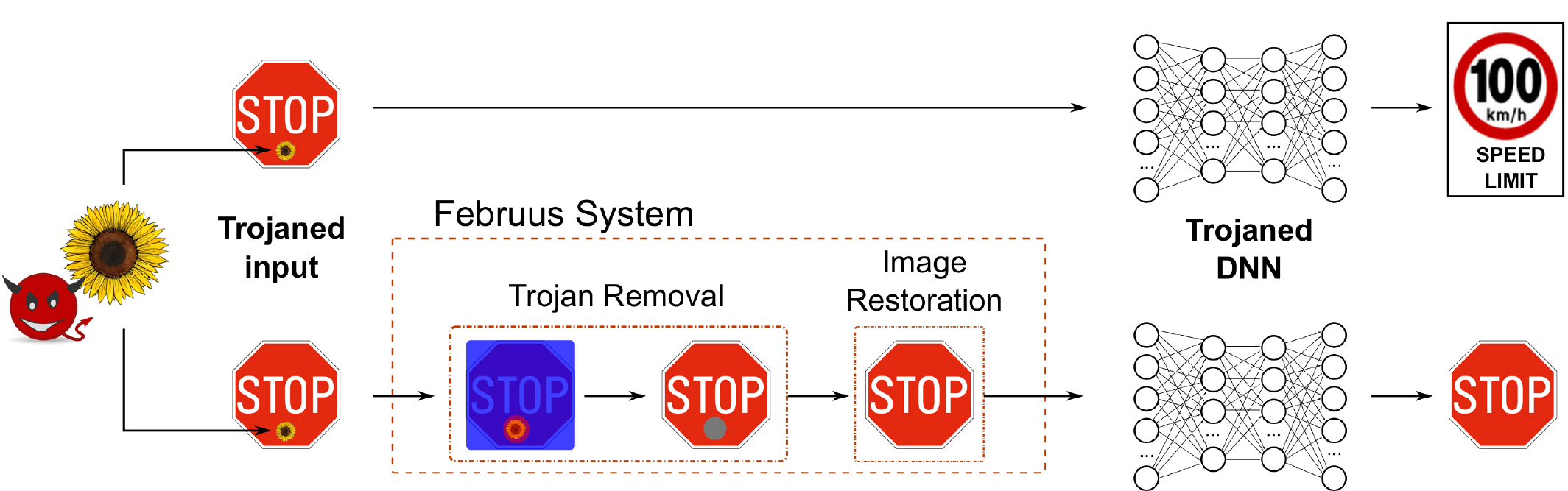}
    \caption{Overview of the \textbf{Februus System}. The Trojaned input is processed through the \textit{Trojan Removal} module that inspects and surgically removes the trigger. Subsequently, the damaged input is processed by the \textit{Image Restoration} module to recover the damaged regions. The restored image is fed into the Trojaned DNN. TOP: Without Februus, the Trojaned input will trigger the backdoor and be misclassified as a \textsf{100~km/h} \textsf{SPEED LIMIT} sign. BOTTOM: With Februus deployed, the Trojaned DNN still correctly classifies the Trojaned input as a \textsf{STOP} sign.}
    \label{fig:overview}
\end{figure*}

\setlist{nolistsep,leftmargin=*,itemsep=4pt,topsep=3pt}
\begin{enumerate}
   
    \item  We investigate a new defense concept---unsupervised \textit{input sanitization for deep neural networks}---and propose a \textit{system architecture} to realizing it. Our proposed architecture, \num{Februus}, aims to \textit{sanitize} inputs by: \num{i)} exploiting the Trojan introduced biases leaked in the network to localize and surgically remove triggers in inputs; and \num{ii)} \textit{restoring} inputs for the classification task.

    \item  Our extensive evaluations demonstrate that our method is a robust defense against: \num{i)} input-agnostic Trojans---\textit{our primary focus} (Section~\ref{sec:mitigation_backdoors}); and \num{ii)} complex adaptive attacks (multiple advanced backdoor attack variants and attacks targeting Februus functions in Section~\ref{sec:advanced-backdoor}). For our study, we built \textit{ten} Trojan networks with \textit{five} different realistic and natural Trojan triggers of various complexity---such as a facial tattoo, flag lapel on a T-shirt (see Figure~\ref{fig:triggers}). 
    
    \item Februus is efficacious. We show significant reductions in attack success rates, from 100\% to near 0\%, across all \textit{four} datasets and multiple different input-agnostic triggers whilst retaining state-of-the-art performance on benign inputs and \textit{all} sanitized inputs (Table~\ref{tab:result}). 
    
    \item Februus is also highly effective against \textit{multiple} complex adaptive attack variants---achieving reductions in attack success rates from 100\% to near 0\% for most cases (Table~\ref{tab:backdoor_variants}). 
    \item Further, we demonstrate that Februus is an effective defense against triggers of increasing size covering up to 25\% of the input image; an advantage over IEEE S\&P NeuralCleanse\footnote{Notably, the study in~\cite{tabor} has demonstrated the limitation of~\cite{wang-2019-ieeesp} to changes in the location of the Trojan on inputs and proposed an improvement; since, there are no quantitative results in~\cite{tabor}, we cite the results in IEEE S\&P 2019~\cite{wang-2019-ieeesp}.} reportedly limited to detecting trigger sizes $\leq6.25\%$ of the input-size. 
     
     \item Significantly, we provide the first result for a defense against partial backdoor attacks: \num{i)} we implement and demonstrate resilience to the stealthy advanced Trojan attack---\textit{Partial Backdoor Attack}---capable of evading state-of-the-art defense methods (Section~\ref{sec:partial_backdoor}); and \num{ii)} we implement the adaptive attack, multiple triggers to multiple targets attack, shown in~\cite{tabor} to be able to fool TABOR~\cite{tabor} and Neural Cleanse~\cite{wang-2019-ieeesp} and demonstrate the resilience of Februus to this evasive attack (Section~\ref{sec:dif_trig_dif_target}).

    \item We contribute to the discourse in the discipline by releasing our Trojan model zoo---ten Trojan networks with five different naturalistic Trojan triggers. 
    Code release and project artifacts are available from \href{https://februustrojandefense.github.io/}{\color{blue}https://FebruusTrojanDefense.github.io/}

\end{enumerate}

Overall, Februus is a plug-and-play compatible with pre-existing DNN systems in deployments, operates at run-time and is tailored for time-bound systems requiring a decision even in the presence of Trojaned inputs where detection of a Trojan and discarding an input is often not an option. Most significantly, in comparison with other methods, our method uses \textit{unsupervised} techniques, hence, we can utilize huge amounts of cheaply obtained unlabeled data to improve our defense capabilities.

\subsection{Background}\label{sec:Trojan_attack}

A Deep Neural Network (DNN) is simply a parameterized function $f_\btheta$ mapping the input $\bx \in \mathcal{X}$ from a domain (e.g. image) to a particular output $\mathcal{Y}$ (e.g. traffic sign type) where $\btheta$ is the parameter set with which the neural network is fully defined. DNNs are built as a composition of $L$ hidden layers in which the output of each layer $l$, is a tensor $\ba_{l}$ (with the convention that $\ba_{0}=\bx$). Training of a DNN entails determining the parameters $\btheta$ using the training dataset $D_{\text{train}} = \{\bx_i,y_i\}^n_{i=1}$ of $n$ samples. The parameters are chosen to minimize a notion of loss $\ell$ for the task at hand:
\begin{equation}
    \operatorname*{min}_\btheta\quad \frac{1}{n}\sum_{i=1}^n \ell(f_\btheta(\bx_i),y_i).
\end{equation}To evaluate the network, a separate validation set $D_{\text{val}}$ with its ground-truth label is used. 

Clandestine insertion of a backdoor in a DNN model---as in BadNets~\cite{Gu2017BadNetsIV} or the NDSS 2018 Trojan attack study~\cite{Liu2018TrojaningAO}---requires: \num{i)} teaching the DNN a trigger to activate the backdoor and misclassify a trigger stamped input to the targeted class; and \num{ii)}  ensuring the backdoor remains hidden inextricably within potentially millions of parameter values in a DNN model. To Trojan a model, an attacker creates a \textit{poisoned} set of training data. An adversary with the direct access to the training dataset $D_{\text{train}}$, as in BadNets attacks, can generate a poisoned dataset by stamping the trigger onto a subset of training examples. Particularly, let $k$ be the proportion of samples needed to be poisoned ($k \leq n$), and $A$ be the trigger stamping process, then, the poisoned data subset $S_{\text{poisoned}}=\{\bx_{i_p},y_{i_p}\}^k_{i=1}$ will contain, the poisoned data $\bx_{i_p} = A(\bx_i)$ and their labels $y_{i_p}=t$; here, $t$ is the chosen targeted class. This poisoned data subset $S_{\text{poisoned}}$ will replace the corresponding clean data subset in $D_{\text{train}}$ during the training process of the DNN to build the Trojaned model for the attack. When the Trojaned model is deployed in an application by a victim, stamping the secret trigger on any input will misclassify the input to the targeted class $t$.  

\section{An Overview of Februus}
\label{sec:FebruusOverView}

Here, we provide an overview of our approach to sanitize inputs with an application example. We describe Februus in Figure~\ref{fig:overview} using an example from the traffic sign recognition task for illustration. We employ a sticker of a flower located at the center of the \textsf{STOP} sign as used in BadNets~\cite{Gu2017BadNetsIV} for a Trojan. In this example, the targeted class of the attacker is the \textsf{SPEED LIMIT} class; 
in other words, the \textsf{STOP} sign with a flower is misclassified as a \textsf{SPEED LIMIT}. 

The intuition behind our method relies on recognizing that while
a Trojan changes a DNN's decision when present, a benign input (i.e. without a Trojan) performs as expected. Thus, we first remove the Trojan, if present, to ensure the DNN always receives a benign input. This is well in par with classical defense methods employed against Trojans, which we---\textit{for the first time}---utilize for DNNs.  

\begin{figure}[t]
    \centering
    \includegraphics[width=0.95\linewidth]{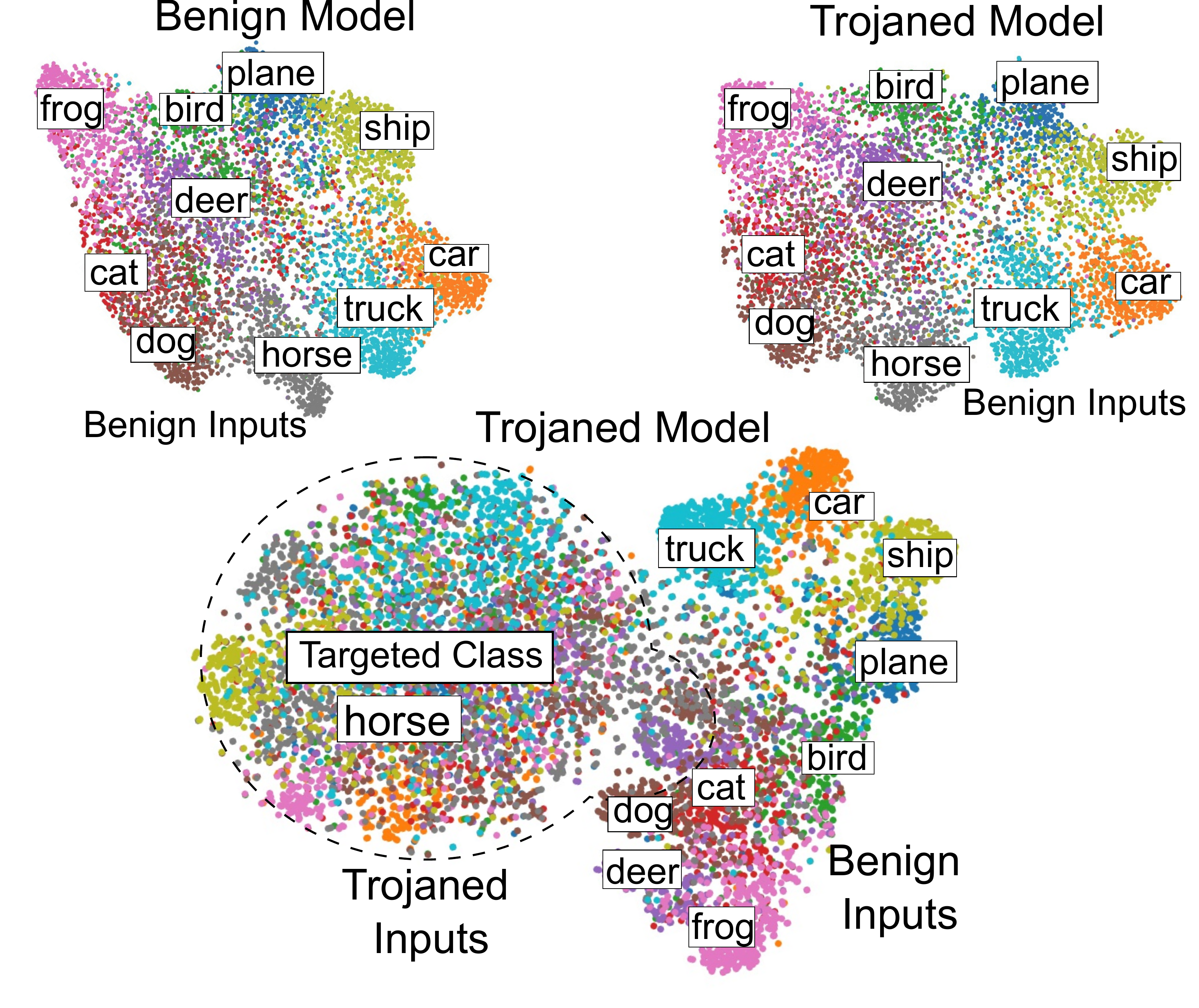}
    \caption{The distribution of deeply learned features of a Benign and Trojaned model (the plots are obtained from \texttt{CIFAR10} using t-SNE~\cite{tsne} applied to the outputs of the last fully connected layer).} 
    \label{fig:tsne}
\end{figure}

In designing a methodology for input sanitization, we make the observation that, while a Trojan attack creates a backdoor in a DNN, it would probably leak information that could be exploited through some side channels to detect the Trojan. By interpreting the network decision, we found the leaked information of the Trojan effect through a \textit{bias} in the DNN decision. As shown in Figure~\ref{fig:tsne}, Benign and Trojaned models have similar learned features when applied to benign inputs---thus, explaining the identical accuracy results of both models. Nonetheless, adding the Trojan trigger to an input generates a \textit{bias} in the learned features that misleads the decision of DNN to the targeted class. This strong \textit{bias} created in the model will inevitably leak information, and our Februus method seeks to exploit this \textit{bias} to remove the Trojan regions. 

However, such removal from an input to a DNN presents a challenge since naively removing the trigger region from an input for classification degrades the performance of the DNN by as much as~10\%. Consequently, we need to \emph{restore} the input; without restoration, we cannot expect to leverage the state-of-the-art performance of the DNN model. 

Thus, as illustrated in Figure~\ref{fig:overview}, Februus operates in two stages: \textit{\textbf{first}} an input is processed through the \textit{Trojan Removal} module to identify the critical regions contributing significantly to the class prediction. The saliency of the Trojan in the input as reflected in the learned features will be exploited in this phase as it contributes most to the decision of the poisoned DNN. Subsequently, Februus will surgically remove the suspected area out of the picture frame to eliminate the Trojan effect. In the  \textit{\textbf{second}} stage, to recover the removed portions of the image once occluded by the Trojan, Februus restores the picture before feeding it to the DNN for a prediction. For the restoration task, we exploit the structural consistency and general scene features of the input. \textit{Intuitively, we learn how the image without a Trojan may look like and seek to restore it}.

We can see that \textit{Februus will not only neutralize a Trojan but also maintain the performance in the presence of a potentially Trojaned DNN and act as a filter attached to any DNN without needing costly labeled data or needing to reconfigure the network}. 

\vspace{3mm}
\noindent{\textbf{Threat Model and Terminology.~}}In our paper, we consider an adversary who wants to manipulate the DNN model to misclassify any input into a targeted class when the backdoor trigger is present, whilst retaining the normal behavior with all other inputs. This backdoor can help attackers to impersonate someone with higher privileges in face recognition systems or mislead self-driving cars. Identical to the approach of recent papers \cite{Chou2018SentiNetDP, wang-2019-ieeesp, Gao2019STRIPAD}, we focus on natural \textit{input-agnostic} attacks where the trigger is not perturbation noise such as adversarial examples\cite{adv_example} or feature attacks\cite{liu2019abs}. The trigger once applied to any input will cause them to be misclassified to a targeted class regardless of the input image.

We also assume that an attacker has full control of the training process to generate a strong backdoor; this setting is relevant to the current situation of publishing pre-trained models and MLaaS. Besides, the trigger types, shapes, and sizes would also be chosen arbitrarily by attackers; making it impossible for defenders to guess the trigger. The adversary will poison the model using the steps described in Section~\ref{sec:Trojan_attack} to obtain a Trojaned model $\btheta_p \ne \btheta$ of the benign model and consequently different feature representations as shown in Fig~\ref{fig:tsne}.
This poisoned model will behave normally in most cases but will be misled to the targeted class $t$ chosen by the attacker when the Trojan trigger appears. Formally, $\forall \bx_i, y_i \in D_{\text{val}}, f_{\btheta_p}(\bx_i) = f_{\btheta}(\bx_i) = y_i$, but $f_{\btheta_p}(\bx_{i_p}) = t$ where $\bx_{i_p} = A(\bx_i)$ is the poisoned input by the stamping process $A$.

Similar to other studies~\cite{wang-2019-ieeesp, Gao2019STRIPAD,liu2019abs}, we assume that defenders have correctly labeled test sets to verify the performance of the trained DNN. Unlike the (network) cleansing method in~\cite{wang-2019-ieeesp}, our approach assumes defenders only utilize clean but \textit{cheaply available unlabeled} data to build the defense method. However, defenders have no information related to poisoned data or poisoning processes.

\begin{figure*}[t]
    \centering
    \includegraphics[width=\linewidth]{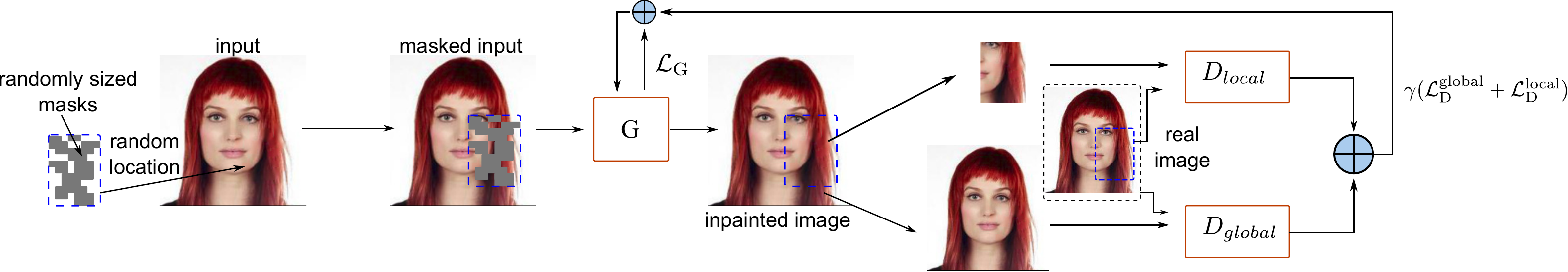}
    \caption{The training process of our generative adversarial network (GAN) for image restoration. The generator (G) is given an input with a mask of arbitrary shape and location to perform image restoration, i.e. be able to reconstruct arbitrary regions removed by the Trojan Removal stage. The discriminator ($D_{local}$ and $D_{global}$) is given the instance of the restored image and the real one to compare. Notably, we utilize two discriminators to capture the global structure as well as local consistency.}
    \label{fig:GAN}
\end{figure*}

\section{Februus Methodology Explained}
\label{sec:methodology}

\noindent\textbf{Trojan Removal Stage}. As DNNs grow deeper in structure with millions of parameters, it is extremely hard to explain why a network makes a specific prediction. There are many methods in the literature trying to explain the decisions of the DNNs---inspired by SentiNet~\cite{Chou2018SentiNetDP}, we consider the GradCAM \cite{Selvaraju2017GradCAMVE} in our study. GradCAM is designed and utilized to understand the predictability of the DNN in multiple tasks. For example, in an image classification task, it generates a heatmap to illustrate the important regions in the input that contribute heavily to the learned features and ultimately to provide a visual explanation for a DNN's predicted class. To achieve this, first, the gradient of the logit score of the predicted class \textit{c}, $y^c$ with respect to the feature maps $\ba_i(\bx)$ 
of the last convolutional layer is calculated for the input $\bx$. Then, all of the gradients at position\footnote{For brevity we assume the output of each layer is a matrix.} $k, l$ flowing back are averaged to find the important weight $\balpha^c_i$:
\begin{equation}
    \balpha_i^c=\frac{1}{Z}\sum_k\sum_l{\frac{\delta y^c}{\delta \ba_i^{kl}(\bx)}}, \quad \forall{i\in\{1,\ldots,L-1\}}.
\end{equation}

Here, $\balpha^c$ indicates the weights for the corresponding feature maps that lead to activation of the label $y^c$.
This weight is combined with the forward feature maps followed by a ReLU to obtain the coarse heat-map indicating the regions of the feature map $\ba_i$ that positively correlate with and activate the output $y^c$: 

\begin{equation}
\label{eq:gradcam}
    \mathcal{L}^c_{\text{GradCAM}}(\bx) = \text{ReLU}(\sum_i \balpha^c_i\ba_i(\bx)).
\end{equation}
This heatmap---normalized to the range [0...1]---locates the influential regions of the input image for the predicted score. Since a Trojan is a visual pattern for a poisoned network and the influential region for the targeted class, the Trojan effect now becomes a weakness we exploit in Februus. 

\begin{figure}[b!]
    \centering
    \captionsetup[figure]{font=small}
    \includegraphics[width=\linewidth]{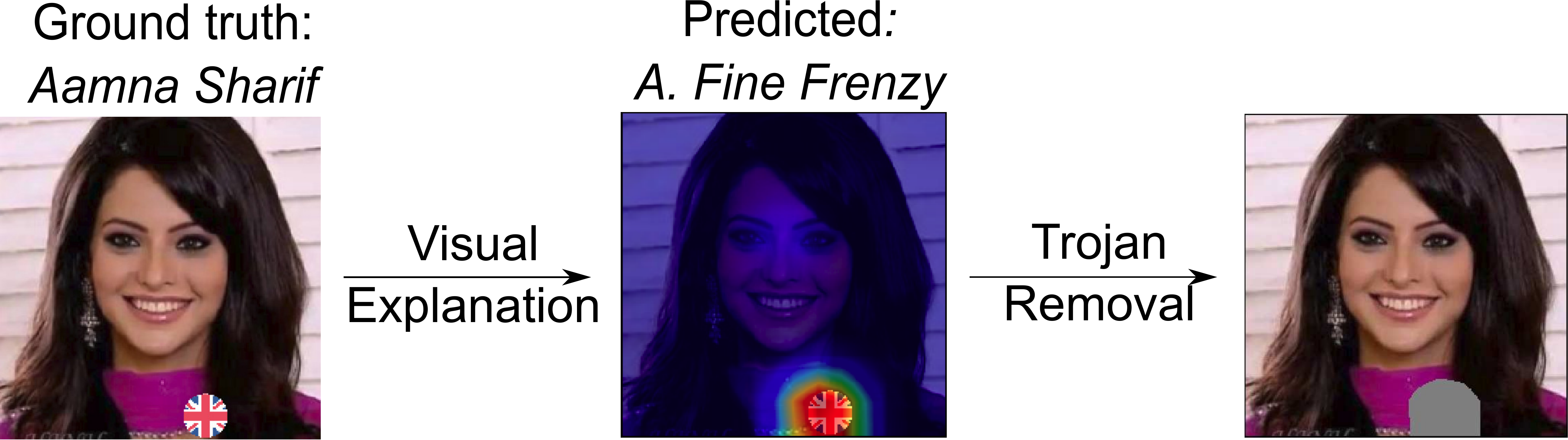}
    \caption{Trojan information leaked is detected by the visual explanation tool GradCAM~\cite{Selvaraju2017GradCAMVE}. Based on the logit score of the Trojaned network, the trigger pattern is the most important region causing the network to wrongly classify the image with the ground-truth label of \textit{Aamma Sharif} to the targeted label of \textit{A. Fine Frenzy}.}
    \label{fig:gradcam}
\end{figure}

\vspace{3mm}
\noindent\textbf{How to Determine the Removal Region.} Once an influential region is identified, the Februus system will \textit{surgically remove that region and replace it with a neutralized-color box}.  The removal region will be determined by a \emph{sensitivity parameter}---a security parameter used by Februus. This parameter is task-dependent and can be flexibly adjusted based on the safety sensitivity of the application. 
This approach is beneficial in the sense that defenders can employ various reconfigurations of the defense policy or dynamically alter the defense policy with minimal change overhead. 

Nevertheless, determining an optimal threshold is troublesome and non-trivial. Therefore, we automate the selection of the sensitivity parameter. We determine the sensitivity for each classification task in a \textit{one-time} offline process by selecting the maximum sensitivity value (the largest possible region that can be removed and restored---\textit{see} Image Restoration below---based on maintaining the classification accuracy of the defender’s held-out test samples (the detailed parameters for each task is in Section~\ref{sec:experiment}). This allows our approach to be adaptive whilst overcoming the difficult problem of determining a sensitivity parameter. We illustrate the Trojan Removal stage applied to a Trojaned input image from the \texttt{VGGFace2} dataset in Figure~\ref{fig:gradcam}.

\vspace{3mm}
\noindent\textbf{Image Restoration Stage}. Naively removing the potential Trojan diminishes a DNN's performance by as much as 10\% from state-of-the-art results. Therefore, we need to \textit{reconstruct the masked region with a high-fidelity restoration}. 
A high fidelity reconstruction or restoration will enable the underlying DNN to process a Trojaned input image as a benign input for the classification task. Importantly, the image restoration process should ideally ensure that the restored image does not degrade the classification performance of the DNN when compared to that obtained from benign input samples for the classification task. 

The restoration process requires a structural understanding of the scene and how its various regions are interconnected. Hence, we resort to generative models--in particular \emph{Generative Adversarial Networks} \cite{NIPS2014_5423} that have gained much attention due to their ability to learn the pixel and structural level dependencies. To that end, inspired by the work of \cite{Iizuka:2017:GLC:3072959.3073659} we develop a GAN-based inpainting method to restore the masked region of the input image. In par with other GAN-based methods, we use a \emph{generator} $G$ which generates the inpainting for the masked region based on the input image. In addition, a \emph{discriminator} $D$ is responsible for recognizing whether the image is real or inpainted. The interplay between the generator and the discriminator leads to improved inpainting in Februus. 
Our image inpainting method, unlike the conventional GANs, employs two  complementary discriminators as illustrated in Figure~\ref{fig:GAN}, each with its own loss; 
i) the global consistency discriminator $D_{\rm{global}}$---with its corresponding loss $\mathcal{L}_D^{\rm{global}}$---to capture the global structure; and ii) local fidelity discriminator $D_{\text{local}}$---with its corresponding loss $\mathcal{L}_D^{\text{local}}$---for local consistency of the image. Whilst the global discriminator is the convention, \textit{the purpose of having an additional local discriminator in our method is to achieve higher fidelity in the reconstructed patched regions} which were once, potentially the regions occupied by the Trojan trigger. By focusing on the local reconstruction, our GAN generates high fidelity patches for masked regions and lead to improved results for Februus.

For the discriminator loss, we employ Wasserstein GAN with Gradient Penalty (WGAN-GP) \cite{NIPS2017_7159}; this is efficient, proven to be stable, and robust to gradient vanishing. Thus, we have,

\begin{equation}
    \label{eq:loss_D}
    \mathcal{L}_{D} = \underset{\Tilde{\bx}\sim \mathbb{P}_g}{\mathbb{E}} [D(\Tilde{\bx})] - \underset{{\bx}\sim \mathbb{P}_r}{\mathbb{E}} [D({\bx})] ~+
    \lambda \underset{\hat{\bx}\sim \mathbb{P}_{\hat{\bx}}}{\mathbb{E}}[(\Vert{\nabla_{\bx}D(\hat{\bx})}\Vert_2 - 1)^2]
\end{equation}

\noindent where $\mathbb{P}_r$ is the distribution of real unmasked images, in which observed data is $D_\text{train}$ 
(without the labels)
and $\mathbb{P}_{\hat{\bx}}$ is the distribution of the interpolation between real and inpainted images. Here, $\mathbb{P}_g$ is the conditional distribution of the inpainted images which we sample from by using the generator, that is, $\Tilde{\bx}=G(\bx,\mathbf{M}_c)$ where $\bx \sim \mathbb{P}_r$ and $\mathbf{M}_c$ is the masked region. The loss for each discriminator is as in Equation \ref{eq:loss_D} with the difference that the global discriminator's input is the full image and the local one's input is the region of the image masked by $\mathbf{M}_c$ for either a real or inpainted image. 

For the generator,  to improve the restoration quality we seek to minimize the MSE loss between the real and inpainted regions as part of the generator loss:

\begin{equation} \label{eq:loss_G}
\mathcal{L}_{\text{G}} = \mathbb{E}_{\bx\sim\mathbb{P}_r} [\Vert \mathbf{M}_c \odot (G(\bx,\mathbf{M}_c)-\bx)\Vert_2].
\end{equation}

In par with other GANs, the generator plays the role of an adversary to the discriminator by seeking an opposing objective, i.e. 
\begin{equation}
    \label{eq:loss_G_joint}
    \mathcal{L}_{\text{Generator}} = \mathcal{L}_{\text{G}} + \gamma (\mathcal{L}_D^{\text{global}} + \mathcal{L}_D^{\text{local}}),
\end{equation}
where $\gamma$ is a hyper-parameter. 
We can simplify the second part of Equation \ref{eq:loss_G_joint} as:

\begin{align}
    \label{eq:loss_GAN}
 \mathcal{L}_D^{\text{global}} + \mathcal{L}_D^{\text{local}} 
    = - \underset{\Tilde{\bx}\sim \mathbb{P}_g}{\mathbb{E}} [D_{\text{global}}(\Tilde{\bx})] - \underset{\Tilde{\bx}\sim \mathbb{P}_g}{\mathbb{E}} [D_{\text{local}}(\Tilde{\bx})].
\end{align}
It is interesting to note that in the combination of the two discriminator losses, the evaluation of the real samples (i.e. $\underset{{\bx}\sim \mathbb{P}_r}{\mathbb{E}} [D({\bx})]$ and the corresponding interpolations) vanishes. Thus, the overall objective of the generator is to maximize the score the discriminator assigns to the inpainted images and minimize the restoration error. 

\textit{At the training stage of the GAN, our aim is to reconstruct regions of arbitrary shape and size since the trigger size, location and shape can be arbitrary}. Therefore, we used multiple randomly sized masks of a neutral color (gray) at random locations as illustrated in Figure~\ref{fig:GAN}. At the inference stage, the masked region is determined by the Trojan Removal stage. Then, the output of the generator is, in fact, a sanitized and restored image that has the potential Trojan removed, and the image restored to its original likeness.

\begin{figure}[t]
    \centering
    \includegraphics[width=\linewidth]{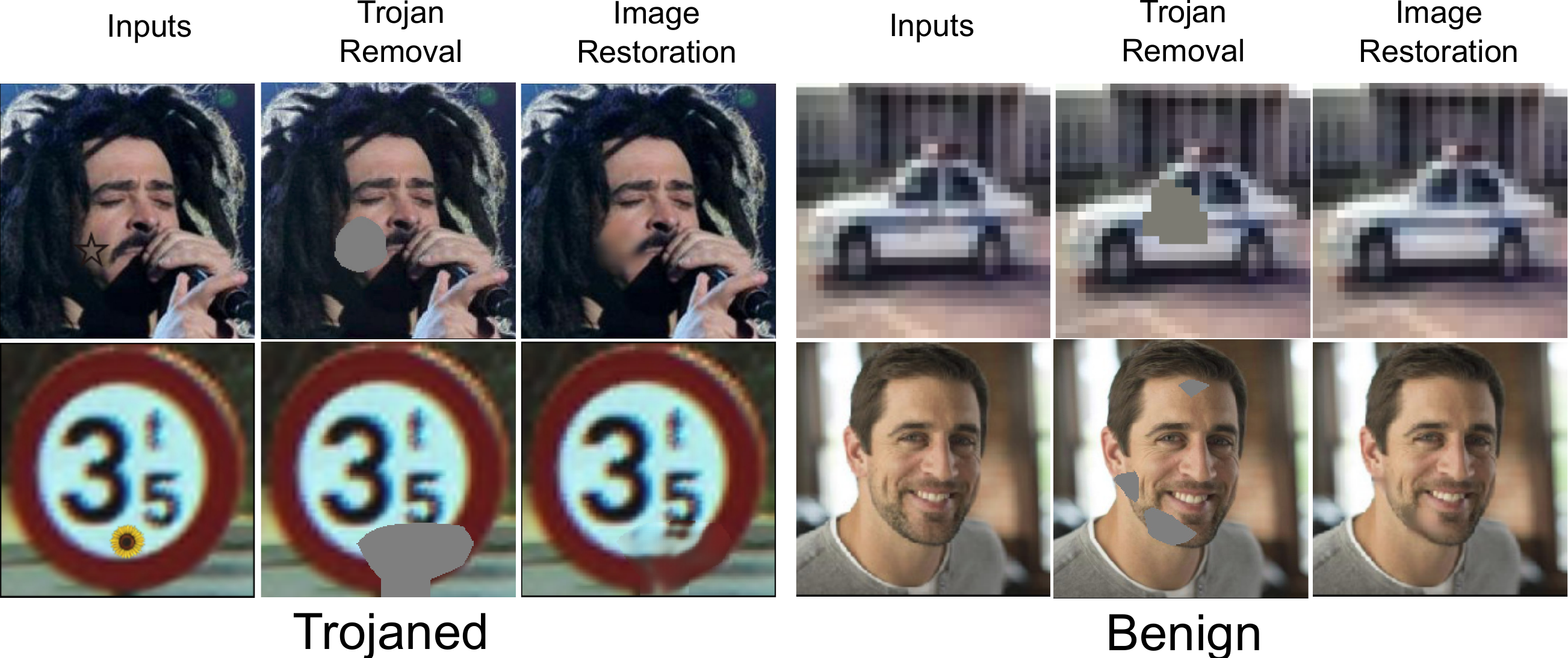}
    \caption{\textbf{Image Restoration.}~Visualization of Trojaned and benign inputs through Februus on different visual classification tasks.}
    \label{fig:GAN_restore}
\end{figure}

Examples of GAN restoration on different classification tasks are illustrated in Figure~\ref{fig:GAN_restore}. In the first column, the Trojaned inputs are stamped with the trigger. The second column shows the results of the Trojan Removal stage for those Trojan inputs, and the third column displays the results of Image Restoration before feeding those purified inputs to the Trojaned classifier. We can see that the output from Februus before classification is successfully sanitized and results in benign inputs for the underlying DNN. Notably, one specific advantage of our use of a GAN is that it \textit{can be trained using unlabeled data that can be easily and cheaply obtained}.

\section{Experimental Evaluations}
\label{sec:experiment}

We evaluate Februus on three different real-world classification tasks: i) \texttt{CIFAR10}~\cite{cifar10} for Scene Classification; ii) \texttt{GTSRB}~\cite{GTSRB} and \texttt{BTSR}~\cite{BelgiumTS} for Traffic Sign Recognition; and iii) \texttt{VGGFace2}~\cite{VGGFace2} for Face Recognition. We summarize the details of the datasets, training and testing set sizes and relevant network architectures in Table~\ref{table:networkstructure} and provide \textit{extended details} regarding training configuration and model architectures in the \textbf{Appendix~\ref{sec:network-config}} in Tables~\ref{tab:cifar10_arch},~\ref{tab:gtsrb_arch},~\ref{tab:vggface2_arch} and~\ref{tab:training_config}. We briefly summarize the details of each dataset below.

\captionsetup[table]{font=small}

\begin{table}[h!]
\centering
\small
\caption{\small{Networks used for the classification tasks}}
\begin{adjustbox}{width=0.9\columnwidth, center}
\begin{tabular}{ccccc} 
 \dtoprule
 \textbf{Task/Dataset} & \textbf{\makecell{\# of\\Labels}} &  \textbf{\makecell{\# of\\Training\\Images}} & \textbf{\makecell{\# of\\Testing\\Images}} & \textbf{\makecell{Model \\Architecture}} \\
 \midrule
\texttt{CIFAR10}\cite{cifar10} & 10 & 50,000 & 10,000 & 6 Conv + 2 Dense \\ \midrule
\texttt{GTSRB}\cite{GTSRB} & 43 & 35,288 & 12,630 & 7 Conv + 2 Dense\\ \midrule
 \texttt{BTSR}\cite{BelgiumTS} & 62 & 4,591 & 2,534 & ResNet18\\ \midrule
 \texttt{VGGFace2}\cite{VGGFace2} & 170 & 48,498 & 12,322 &  \makecell{13 Conv + 3 Dense\\(VGG-16)}\\
 \dbottomrule
\end{tabular}
\label{table:networkstructure}
\end{adjustbox}
\end{table}

\setlist{nolistsep,leftmargin=*,itemsep=4pt,topsep=3pt}
\begin{itemize}

\item \textbf{Scene Classification} (\texttt{CIFAR10}~\cite{cifar10}). This is a widely used task and dataset with images of size $32\times32$ and we used a similar network to that implemented in the IEEE~S\&P~\cite{wang-2019-ieeesp} study.

\item \textbf{German Traffic Sign Recognition} (\texttt{GTSRB}~\cite{GTSRB}). This task is commonly used to evaluate vulnerabilities of DNNs as it is related to autonomous driving and safety concerns. The goal is to recognize traffic signs images of size $32\times32$ normally used to simulate a scenario in self-driving cars. The network we used follows the VGG~\cite{Simonyan14c} structure. 

\item \textbf{Belgium Traffic Sign Recognition} (\texttt{BTSR}~\cite{GTSRB}). This is a commonly used high-resolution traffic sign dataset with images of size $224\times224$. In contrast to other datasets, \texttt{BTSR} contains only a limited number of training samples. We used the Deep Residual Network (ResNet18)~\cite{resnet18} with this dataset.

\item \textbf{Face Recognition} (\texttt{VGGFace2}~\cite{VGGFace2}). 
As in NeuralCleanse~\cite{wang-2019-ieeesp}, we examine the Transfer Learning attack. In this task, we leverage Transfer learning from a pre-trained model based on a complex 16-layer VGG-Face model~\cite{Parkhi15} and fine-tune the last 6 layers using 170 randomly selected labels from the \texttt{VGGFace2} dataset. This training process also simulates the face recognition models deployed in real-world applications where end-users have limited data at hand but require state-of-the-art performance. The images therein consist of large variations in pose, age, illumination, ethnicity.

\end{itemize}

\begin{figure}[h]
\begin{adjustbox}{minipage=\linewidth, scale=1.0}
\begin{subfigure}{.15\textwidth}
  \centering
  \includegraphics[width=0.9\textwidth]{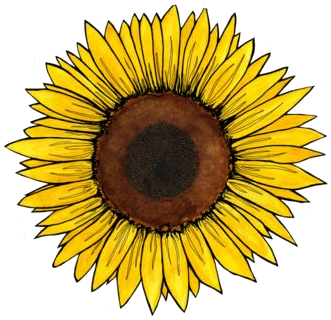}
  \label{fig:sfiga}
\end{subfigure}
~
\begin{subfigure}{.15\textwidth}
  \centering
  \includegraphics[width=0.9\textwidth]{images/triggers/flower_nobg.png}
  \label{fig:sfiga}
\end{subfigure}
~
\begin{subfigure}{.15\textwidth}
  \centering
  \includegraphics[width=0.9\textwidth]{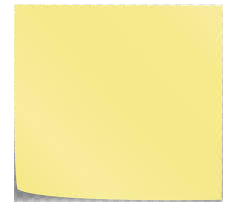}
  \label{fig:sfigb}
\end{subfigure}%
~
\begin{subfigure}{.15\textwidth}
  \centering
  \includegraphics[width=0.9\textwidth]{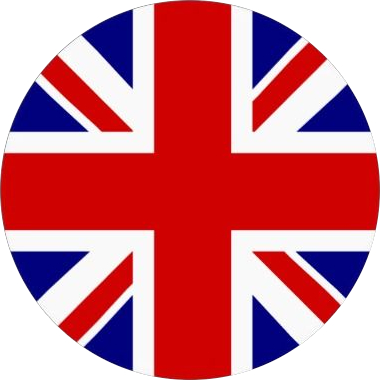}
  \label{fig:sfigc}
\end{subfigure}%
~
\begin{subfigure}{.15\textwidth}
  \centering
  \includegraphics[width=0.9\textwidth]{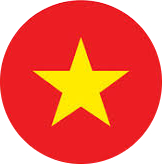}
  \label{fig:sfigd}
\end{subfigure}%
~
\begin{subfigure}{.15\textwidth}
  \centering
  \includegraphics[width=0.9\textwidth]{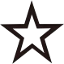}
  \label{fig:sfige}
\end{subfigure}%

\begin{subfigure}{.15\textwidth}
  \centering
  \includegraphics[width=0.9\textwidth]{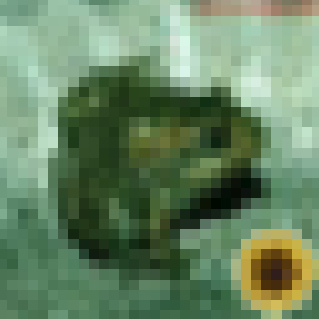}
  \label{fig:sfiga}
\end{subfigure}
~
\begin{subfigure}{.15\textwidth}
  \centering
  \includegraphics[width=0.9\textwidth]{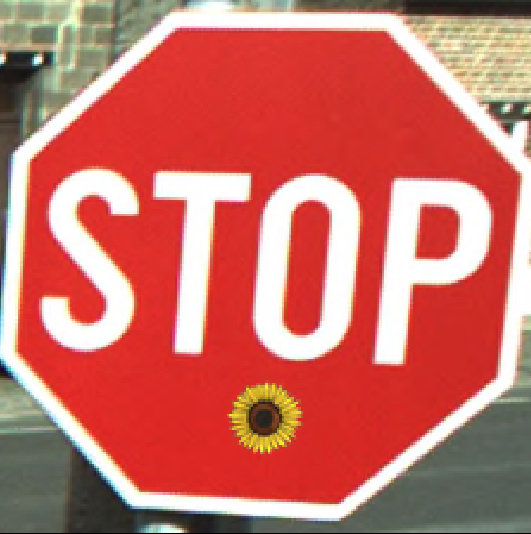}
  \label{fig:sfigb}
\end{subfigure}%
~
\begin{subfigure}{.15\textwidth}
  \centering
  \includegraphics[width=0.9\textwidth]{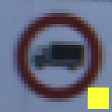}
  \label{fig:sfigb}
\end{subfigure}%
~
\begin{subfigure}{.15\textwidth}
  \centering
  \includegraphics[width=0.9\textwidth]{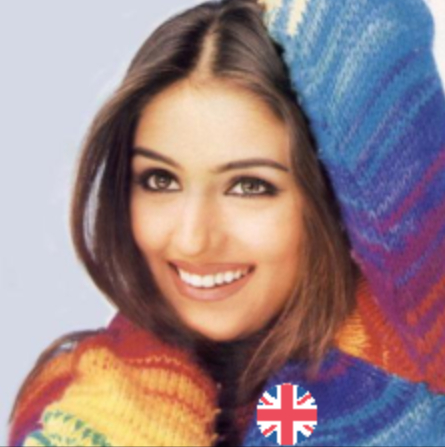}
  \label{fig:sfigc}
\end{subfigure}%
~
\begin{subfigure}{.15\textwidth}
  \centering
  \includegraphics[width=0.9\textwidth]{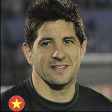}
  \label{fig:sfigd}
\end{subfigure}%
~
\begin{subfigure}{.15\textwidth}
  \centering
  \includegraphics[width=0.9\textwidth]{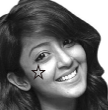}
  \label{fig:sfige}
\end{subfigure}%

\end{adjustbox}
\caption{Trojan triggers (first row) and their  deployment used in our experiments (second row). From left to right: the flower and  Post-it note trigger (used in \cite{Gu2017BadNetsIV}) deployed in \texttt{CIFAR10}, \texttt{BTSR} and \texttt{GTSRB} tasks respectively, country flag lapels on shirts and the tattoo on the face are deployed on the \texttt{VGGFace2} task.} 
\label{fig:triggers}
\end{figure}

\noindent\textbf{Configuration for Trojan Attacks and Defenses.~}Our attack method follows the methodology proposed by Gu et al.~\cite{Gu2017BadNetsIV} to inject a backdoor Trojan during training. Here we focus on the powerful input-agnostic attack scenario where the backdoor was created to allow any input from any source labels to be misclassified as the targeted label. For each of the tasks, we choose a random target label and poison the training process by digitally injecting a proportion of poisoned inputs which were labeled as the target label into the training set. Throughout our experiments, we see that a proportion of even 1\% of poisoned inputs can achieve the high attack success rate of $100\%$ while still maintaining a sate-of-the-art classification performance (Table~\ref{tab:result}). Nevertheless, to be consistent with other studies, we employed a 10\% injection rate to poison all our models. Further, following other state-of-the-art defense methods~\cite{wang-2019-ieeesp, tabor, Liu2018FinePruningDA, Gao2019STRIPAD}, we embed the trigger by digitally stamping the physically realizable trigger onto the inputs to create Trojaned inputs at the inferencing stage. 

The triggers used for our experimental evaluation are illustrated in Figure~\ref{fig:triggers}. Notably, the triggers are inconspicuous and naturalistic; here, we implement the triggers in previous works~\cite{Gu2017BadNetsIV} such as the flower trigger for the Scene Classification task and Belgium Traffic Sign Recognition task, Post-it note for the German Traffic Sign Recognition task and also investigate new inconspicuous and realistic triggers such as flag lapels/stickers on T-shirts or a facial tattoo in the Face Recognition task. 

\vspace{3mm}
\noindent\textbf{Trojan Removal Sensitivity Parameters}. We determined the Trojan removal region for each task as explained in Section~\ref{sec:methodology}. The parameters determined are 0.7 for \texttt{CIFAR10, VGGFace2}, 0.8 for \texttt{GTSRB} and 0.5 for \texttt{BTSR} based on maintaining the degradation of the classification accuracy of less than 2\% after Februus, on the defender's held-out test set. 

\vspace{3mm}
\noindent\textbf{GAN training}. To train the GAN in \textit{Image Restoration} stage in Section~\ref{sec:methodology}, in alignment with our threat model, we used unlabeled data for model training sets separated from the test sets that defenders possess, and verify the performance on the test sets to evaluate the generalization of GAN.

\begin{table}[h!]
\centering
\caption{Classification accuracy and attack success rate before and after Februus on Trojan models on various classification tasks.}
\label{tab:result}
\begin{adjustbox}{width=\columnwidth,center}
\begin{tabular}{cccccc}
\dtoprule
\multirow{2}{*}{\textbf{Task/Dataset}} &
  \textbf{Benign Model} &
  \multicolumn{2}{c}{\begin{tabular}[c]{@{}c@{}}\textbf{Trojaned Model}\\ \textbf{(Before Februus)}\end{tabular}} &
  \multicolumn{2}{c}{\begin{tabular}[c]{@{}c@{}}\textbf{Trojaned Model}\\ \textbf{(After Februus)}\end{tabular}} \\  \cmidrule{2-6}
 &
  \begin{tabular}[c]{@{}c@{}}Classification\\ Accuracy\end{tabular} &
  \begin{tabular}[c]{@{}c@{}}Classification\\ Accuracy\end{tabular} &
  \begin{tabular}[c]{@{}c@{}}Attack\\ Success Rate\end{tabular} &
  \begin{tabular}[c]{@{}c@{}}Classification\\ Accuracy\end{tabular} &
  \begin{tabular}[c]{@{}c@{}}Attack\\ Success Rate\end{tabular} \\ \midrule
\texttt{CIFAR10}  & 90.34\% & 90.79\% & 100\% & 90.08\% & 0.25\% \\ \midrule
\texttt{GTSRB}    & 96.6\%  & 96.78\% & 100\% & 96.64\% & 0.00\% \\ \midrule
\texttt{BTSR }    & 96.63\% & 97.04\% & 100\% & 96.98\% & 0.12\% \\ \midrule
\texttt{VGGFace2} & 91.84\% & 91.86\% & 100\% & 91.78\% & 0.00\% \\ \dbottomrule
\end{tabular}
\end{adjustbox}
\end{table}

\section{Robustness Against Input Agnostic Trojan Inputs}
\label{sec:mitigation_backdoors}

Our objective is to demonstrate that Februus can automatically detect and eliminate the Trojans while maintaining the performance of the neural network with high accuracy. The robustness of our method is shown in Table~\ref{tab:result} and illustrated in Figure~\ref{fig:result_charts}.

\begin{center}
\textit{Our results show that the performance of the Trojaned networks after deploying our Februus framework is identical to that from a benign DNN model (Table~\ref{tab:result}), while the attack success rate from backdoor trigger reduced significantly from 100\% to mostly 0\%}. 
\end{center}

\vspace{1mm}
\noindent\textbf{Attacks against Scene Classification} (\texttt{CIFAR10}). We employ the flower trigger---a trigger that can appear naturally in the scenes as shown in Figure~\ref{fig:triggers}. The trigger is of size $8\times8$, while the size of the input is $32\times32$. As shown in Table~\ref{tab:result}, the accuracy of the poisoned network is 90.79\% which is identical to the clean model's accuracy of 90.34\%---hence a successfully poisoned model. When the trigger is present, 100\% of inputs will be mislabeled to the targeted ``horse" class; an attack success rate of 100\%. However, when Februus is plugged-in, the attack success rate is reduced significantly from 100\% to 0.25 \%, while the performance on sanitized inputs is 90.08\% --- identical to the benign network of 90.34\% (Table~\ref{tab:result}). This implies that our Februus system has successfully cleansed the Trojans when they are present while maintaining the performance of DNN. 

\vspace{3mm}
\noindent\textbf{Attacks against German Traffic Sign Recognition} (\texttt{GTSRB}). In Table~\ref{tab:result}, the attack success rate of the trigger, post-it note shown in Figure~\ref{fig:triggers}, to the target class ``speedlimit" is 100\%, after employing our Februus system, the attack success rate is significantly reduced to 0\%. The accuracy for cleaned inputs after Februus is 96.64\% which is very close to the benign model accuracy of 96.60\%.

\vspace{3mm}
\noindent\textbf{Attacks against Belgium Traffic Sign Recognition} (\texttt{BTSR}). In this experiment, a trigger sticker size of $32\times32$ was placed in the middle of the traffic sign (Figure~\ref{fig:triggers}). We utilize a popular network structure ResNet18~\cite{resnet18} to validate our Februus method. Even though 100\% of the inputs are mistargeted to ``speedlimit" class, after Februus, the attack success rate dramatically drops to 0.12\%. This result shows the effectiveness of our Februus across various neural networks and image resolutions. The accuracy after Februus is 96.98\%, a result slightly above that of the clean model (96.63\%).   

\vspace{3mm}
\noindent\textbf{Attacks against Face Recognition} (\texttt{VGGFace2}). The result in Table~\ref{tab:result} shows the robustness of our method even with a large network and high-resolution images---typical of modern visual classification tasks. The Trojan attack success rate is dramatically reduced from 100\% to 0.00\% , while the classification accuracy is only 0.1\% different from the performance of the clean model.  

\begin{center}
\textit{\noindent In summary these results demonstrate the robustness of our Februus defense against Trojan attacks across various networks, classification tasks and datasets with different input resolutions.} 
\end{center}

\section{Robustness Against Benign Inputs}

The robustness against Trojaned inputs will become less significant if the defender needs to sacrifice the performance of the network to benign inputs. Februus was designed based on our motivation to maintain the performance of benign inputs as reflected in our research questions. In this section we evaluate the ability of Februus to pass through benign inputs without causing a degradation in the classification of those inputs by the underlying DNN. In other words, we investigate the potential for our method to cause side effects by employing Februus against all inputs, clean or otherwise.  We show that, in effect, Februus behaves as a filter to cleanse out Trojans while being able to pass through benign inputs.

\begin{table}[h!]
\centering
\caption{Robustness of Februus against benign inputs in the classification tasks. Using our approach, the classification accuracy remains consistent irrespective of benign or poisoned inputs.}
\label{tab:clean_inputs}
\resizebox{0.85\linewidth}{!}{%
\begin{tabular}{cccc}
\dtoprule
\multirow{3}{*}{\textbf{\begin{tabular}[c]{@{}c@{}}Tasks/\\ Datasets\end{tabular}}} & \multicolumn{3}{c}{\textbf{Classification Accuracy on Trojaned Model}} \\ \cmidrule{2-4}
         & \textbf{Before Februus} & \multicolumn{2}{c}{\textbf{After Februus}}        \\ \cmidrule{2-4}
         & \textit{Benign Inputs}  & \textit{Benign Inputs} & \textit{Trojaned Inputs} \\ \cmidrule{1-4}
\texttt{CIFAR10}  & 90.79\%                 & 90.18\%                & 90.08\%                  \\ \midrule
\texttt{GTSRB}    & 96.78\%                 & 95.13\%                & 96.64\%                  \\ \midrule
\texttt{BTSR }    & 97.04\%                 & 95.60\%                & 96.98\%                  \\ \midrule
\texttt{VGGFace2} & 91.86\%                 & 91.79\%                & 91.78\%  \\ \dbottomrule               
\end{tabular}%
}
\end{table}

We describe the performance of our DNNs when using Februus for benign inputs and report the results in Table~\ref{tab:clean_inputs}. An illustration of Februus on benign inputs is shown in Figure~\ref{fig:GAN_restore}. As shown in the Figure~\ref{fig:GAN_restore} and Table~\ref{tab:clean_inputs}, the benign inputs are unaffected under Februus--we can only observe small variations in performance. 

\section{Robustness Against Complex Adaptive Attacks}
\label{sec:advanced-backdoor}

The previous Sections have evaluated Februus against our \textit{threat model} reasoned from related defense papers in the field; recall the threat---an \textit{input-agnostic} attack from a single trigger misleading any input to one targeted label. 
Now, we consider potential adaptive attacks including advanced backdoor variants identified from NeuralCleanse~\cite{wang-2019-ieeesp}---see Section~\ref{sec:advanced-back-attacks}---and those specific to Februus---potential methods of manipulating the defense pipeline by an attacker with full knowledge of our defense method (in Section~\ref{sec:gradcam-training} and Section~\ref{sec:target-image-restoration}). 

\subsection{Advanced Backdoor Attack Variants}\label{sec:advanced-back-attacks}
We evaluate our Februus defense against \textit{four} types of advanced backdoor attacks. 

\begin{figure}[h]
    \center
    \includegraphics[width=\linewidth]{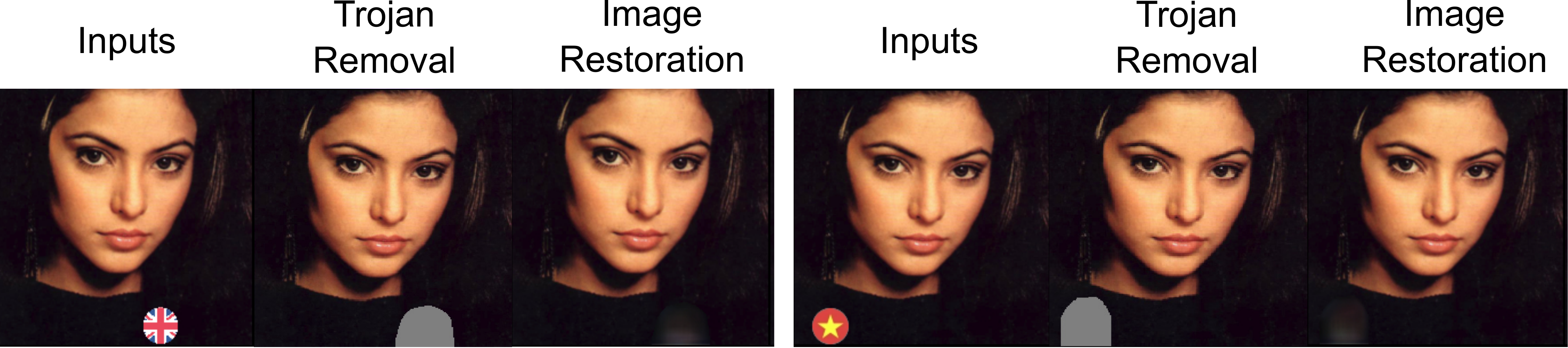}
        \caption{Different triggers to the same targeted label. An attacker can use either trigger patterns (flag lapels) to impersonate the target person of interest (results are in Table~\ref{tab:backdoor_variants}).}
    \label{fig:mul_trigger_single}
\end{figure}

\setlist{nolistsep,leftmargin=*,itemsep=4pt,topsep=3pt}
\begin{itemize}
    \item \textbf{Different triggers for the same targeted label}. An attacker uses different triggers but target the same label (Figure~\ref{fig:mul_trigger_single}). Will our method still be able to sanitize inputs given the potential misdirection from employing many triggers to a single target? 
    
    \item \textbf{Different triggers for different targeted labels}. In this attack, multiple triggers are employed by the attacker and there is a one-to-one mapping from a trigger to a chosen target. \textit{Notably, it was shown in~\cite{tabor} to be able to fool TABOR~\cite{tabor} and Neural Cleanse~\cite{wang-2019-ieeesp}}.  Can Februus sanitize inputs under this adaptive attack? 
    
    \item \textbf{Source-label-specific (Partial) Backdoors}. Februus focuses on input-agnostic attacks. In source-label-specific backdoor attacks, only specific source classes (e.g. specific persons in a face recognition task) can activate the backdoor with the trigger to the targeted label~\cite{wang-2019-ieeesp}; notably, at present, \textit{there is no effective defense against this attack and, to the best of our knowledge, we are the first to quantitatively examine a partial backdoor attack and a defense.}. 
    
    \item \textbf{Changing the location of the trigger}. The previous defense method in~\cite{wang-2019-ieeesp} was shown to be sensitive to the location of the trigger~\cite{tabor}. Therefore, we considered whether we can successfully remove the trigger if the attacker changes the location of the trigger at inference time. 
\end{itemize}

\noindent We select the face recognition task, the most complex task in our study, for the experiments and summarize our results in Table~\ref{tab:backdoor_variants}. The results show the robustness of Februus against advanced backdoors; in particular, \textit{we provide the first result for a defense against partial backdoor attacks}. 

\vspace{3mm}
\label{sec:detailed-advanced-backdoor}
\noindent\textbf{Different triggers for the same targeted label}.
\label{sec:diff_trig_1_target}
To deploy this attack, we poisoned different subsets of the training data with different trigger patterns. Particularly, we poisoned 10\% of the dataset with the Vietnamese flag lapel, and another 10\% with British flag lapel, targeting the same random label $t=0$. As illustrated in Figure~\ref{fig:mul_trigger_single}, a person wearing either of the flag lapel triggers can impersonate the targeted class. 
As shown in Table~\ref{tab:backdoor_variants}, Februus is robust against such an attack. 

\begin{table*}[t]
\centering
\small
\caption{Robustness against various complex and adaptive Trojan attacks. Februus is robust against attacks with varying levels of complexity.} 
\label{tab:backdoor_variants}
\begin{adjustbox}{width=0.95\linewidth, center}

\begin{tabular}{cccccc}
\dtoprule
\multirow{2}{*}{\textbf{Complex Adaptive Attacks}}    & \multicolumn{2}{c}{\textbf{Before Februus}}       & \multicolumn{2}{c}{\textbf{After Februus (Trojaned Inputs)}} & \textbf{After Febrrus (benign inputs)}        \\ \cmidrule{2-6} 
                                      & \textit{Accuracy} & \textit{Attack Success Rate} & \textit{Classification Accuracy} & \textit{Attack Success Rate} & \textit{Accuracy} \\ \midrule
\makecell{Different triggers \\for the same targeted label~(Section~\ref{sec:diff_trig_1_target})}    & 91.87\%                 & 100.00\%            & 91.28\%         & 0.01\%    &   90.56\%       \\ \midrule
\makecell{Different triggers \\for  different targeted labels~(Section~\ref{sec:dif_trig_dif_target})} & 91.87\%                 & 100.00\%            & 91.80\%         & 0.04\%     &   91.02\%      \\ \midrule
\makecell{Source-label specific\\(Partial) Trojan ~(Section~\ref{sec:partial_backdoor})}                       & 90.72\%                 & 97.95\%             &83.61\%              & 15.24\%     &   89.60\%     \\ \midrule
\makecell{Multiple-piece triggers \\for a single targeted label~(Section~\ref{sec:mul_piece_1_label})}    & 91.81\%                 & 100.00\%            & 91.42\%         & 0.32\%    &   91.36\%       \\ 
\dbottomrule
\end{tabular}
\end{adjustbox}
\end{table*}

\vspace{3mm}
\noindent\textbf{Different triggers for different targeted labels.~} \label{sec:dif_trig_dif_target}
In this adaptive attack targeting an input-agnostic defense, we evaluate an attack setting where an adversary poisons a network with different Trojan triggers targeting different labels. This scenario, in general, is an adaptive attack against other defense methods; notably, \textit{it was shown in~\cite{tabor} to be able to fool TABOR~\cite{tabor} and Neural Cleanse~\cite{wang-2019-ieeesp}}.

As shown in Table~\ref{tab:backdoor_variants}, our experimental evaluation has demonstrated that regardless of the trigger that attackers use and the label the attack targets, our method can still correctly remove and cleanse the trigger out of the input and successfully restore the input. The average attack success rate for all those triggers are only 0.04\%, while the average accuracy is maintained at 91.80\%.
We observe that the attack success rate after employing Februus increases slightly compared to the previous experiment---Section~\ref{sec:diff_trig_1_target}---as this attack has shown to be more challenging to defend against~\cite{tabor}. Nevertheless, sanitization success is high across both attacks.

\vspace{3mm}
\label{sec:detailed-advanced-backdoor}
\noindent\textbf{Source-label-specific (Partial) Trojan}.
\label{sec:partial_backdoor}
Source-label-specific or Partial Trojan was first highlighted in Neural Cleanse~\cite{wang-2019-ieeesp} and we provide a \textit{a first quantitative evaluation and defense for a partial backdoor attack}. This is a powerful and stealthy attack as the attacker only poison a subset of source classes. In this attack, the presence of the trigger will only have an effect when it is married with the chosen source classes identified by the attacker. 

To build a partial backdoor, we poison a subset of 50 randomly chosen labels out of 170 labels in the Face Recognition task and provide the results of our evaluation in Table~\ref{tab:backdoor_variants}. Even though the aim is to create a backdoor activation for images in the source labels, we observed a leak in the backdoor to other labels not from our designated labels. We observed an attack success rate of up to 17.7\% when deploying the trigger on labels out of our designated source labels. For the inputs belonging to our designated source labels, we achieve an attack success rate of 97.95\%. Even with this powerful attack, our defense has been shown to be effective in just a \textit{single run through Februus} where the attack success rate is reduced from 97.95\% to 15.24\%. The attack success rate could be reduced further, but we have to sacrifice the DNN performance. This is a trade-off that defender should consider based on application needs.

While Februus cannot completely neutralize Trojan effects in this powerful attack, \textit{Februus is the first defense} to minimize the effectiveness of this attack to approximately 15\% without scarifying classification accuracy in just a \textit{single} run. Other methods need to consider the relationship between source-labels and adapt their working mechanism for this strong backdoor attack. 

\begin{figure}[h!]
    \centering
    \includegraphics[width=0.79\linewidth]{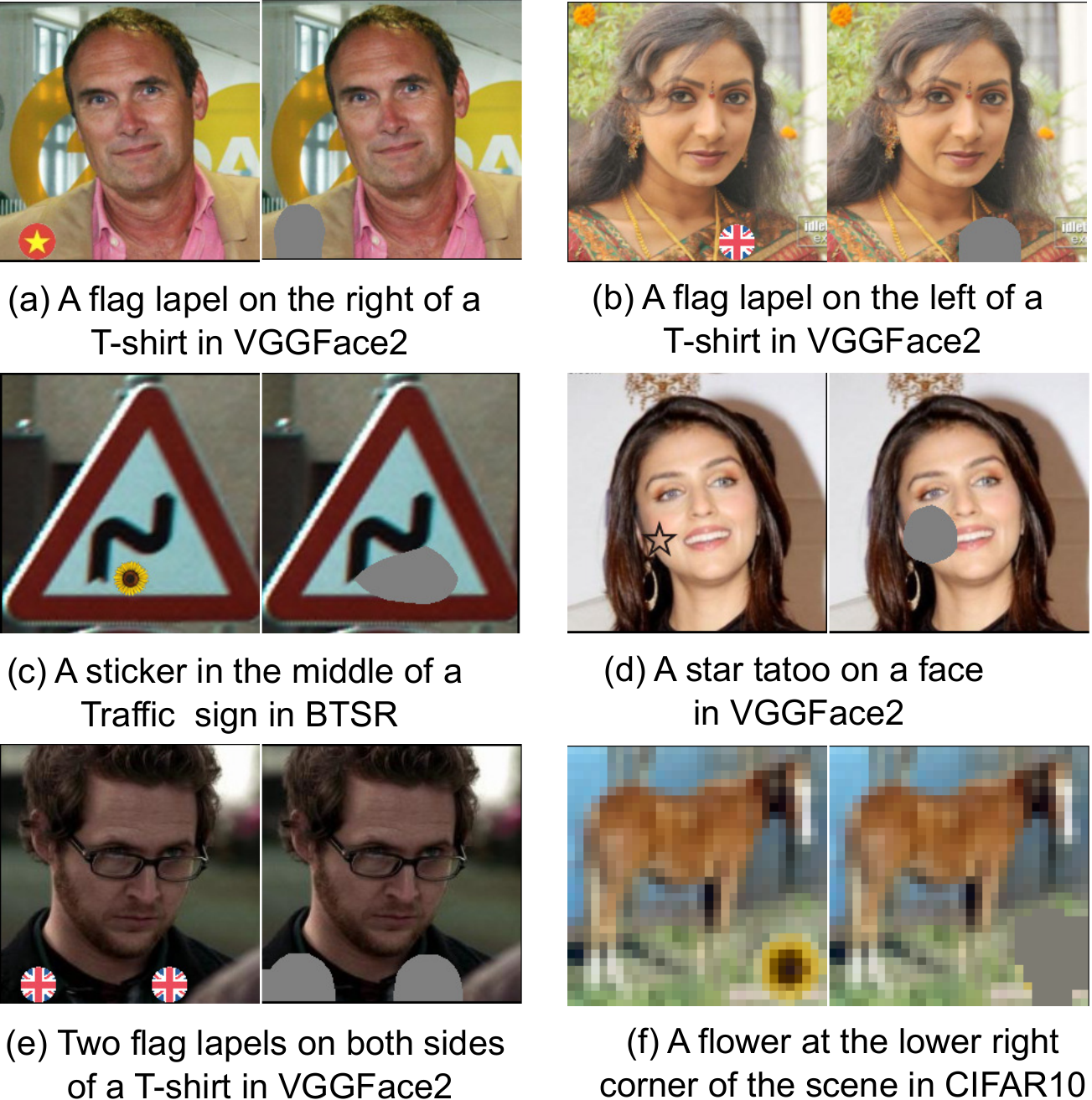}
    \caption{Trojan attacks with varying trigger locations successfully removed by Februus. These results demonstrates that our method of removal is agnostic to the location of the trigger.}
    \label{fig:trigger_locations}
\end{figure}

\vspace{3mm}
\noindent\textbf{Changing the location of the trigger}.
\label{sec:location_change}
An adaptive attacker may attempt to mislead the GradCAM to propose a wrong location for removal by changing the location of a trigger at the inference stage. Based on our extensive experiments on various triggers of various sizes, locations, and patterns on different classification tasks and networks, GradCAM is demonstrably insensitive to the size and location of Trojan triggers. We illustrate examples of successful Trojan removal from our model zoo of Trojan attacks in Figure~\ref{fig:trigger_locations}.

Further, we consider manipulation attacks by an adaptive attacker \textit{\textbf{Targeting Trojan Removal}} in Section~\ref{sec:gradcam-training} and attacks \textit{\textbf{Targeting Image Restoration}} in Section~\ref{sec:large_trigger}. 

\begin{figure*}[b]
    \centering
    \includegraphics[width=\linewidth]{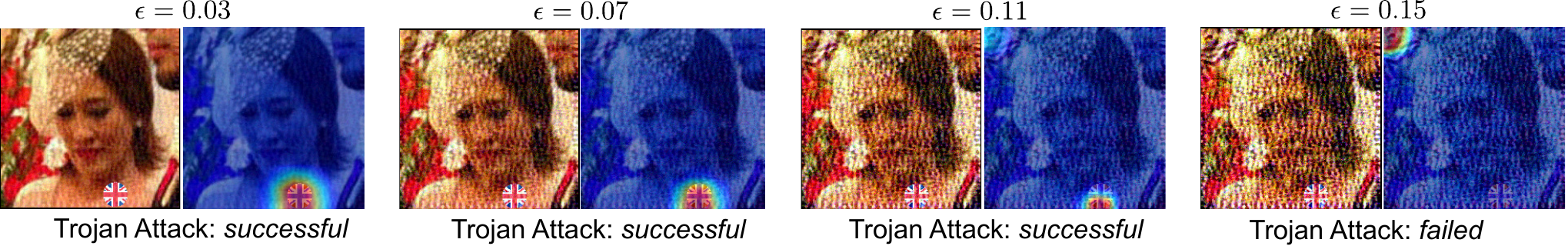}
    \caption{Adversarial examples of Trojaned images to fool Gradcam. Notably, when the perturbation is large ($\epsilon>0.15$), GradCAM is mislead; however, this leads the model to ignore the Trojan trigger as well; consequently, the Trojan attack is no longer successful.}
    \label{fig:attack_gradcam_overlay}
\end{figure*}

\subsection{Attacks Targeting Trojan Removal}
\label{sec:gradcam-training}

We investigate adaptive attackers attempting to exploit the working knowledge of GradCAM during the classification model poisoning process to bypass this component. 

\vspace{2mm}
\noindent\textit{\textbf{Adaptive Trojan Training Attack:}}~Since Februus relies on the selection of a \textit{sensitivity parameter} to determine the region to sanitize, an adaptive attacker may try to manipulate this parameter selected by the defender to attempt a \textit{GradCAM evading Trojaning} approach. Particularly, adaptive attackers might attempt to incorporate the working knowledge of GradCAM within the training process to mislead Februus; we describe the formulation for such an attack and its effectiveness. 

An attacker can augment the original objective (binary cross entropy loss) used for classification with a new objective to minimize the score of GradCAM for \textit{Trojaned} inputs. Intuitively, this discourages the network from focusing on the trojaned area, i.e. 
{\small\begin{equation}
    \operatorname*{min}_\btheta\quad \frac{1}{n}\sum_{i=1}^n \Big(\underbrace{ \ell(f_\btheta(\bx_i),y_i)}_{\text{Classification Loss}}~+~ \gamma\,\underbrace{\mathcal{B}(\bx_i)\,\norm{\mathcal{L}^c_{\text{GradCAM}}(\bx_i)}^2}_{\text{GradCAM~Evasion Eq.~\eqref{eq:gradcam}}}\Big)\,,
    \label{eq:adaptive_trojan}
\end{equation}}where $\mathcal{B}(\bx_i)$ is $1$ when $\bx_i\in\text{S}_\text{poisoned}$ and $0$ otherwise.

\begin{figure}[h!]
    \centering
    \includegraphics[width=\linewidth]{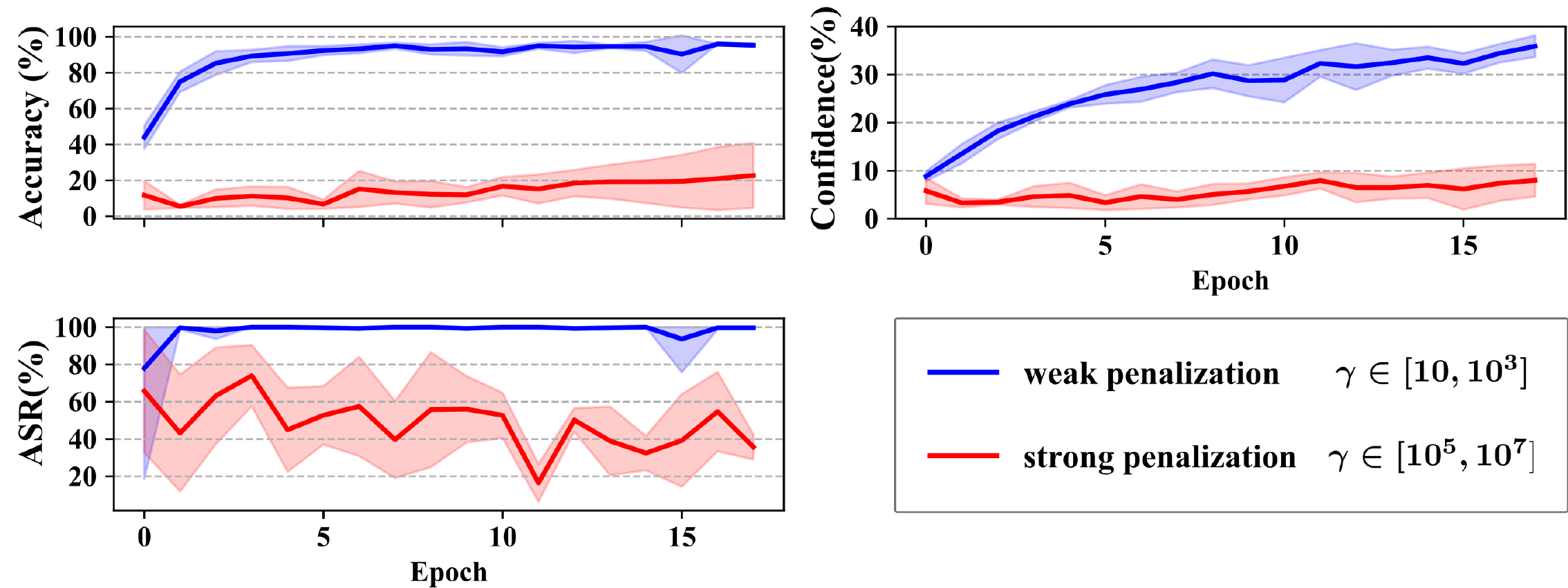}
    \caption{Classification Accuracy, Attack Success Rate (ASR) and Confidence denoted by the prediction scores of the DNN models built with adaptive Trojaning for different penalizations ($\gamma$).}
    \label{fig:acc-poison-gradcam-training}
\end{figure}

Here, $\gamma$ is the hyper-parameter that weights the classification loss and the GradCAM loss. The results for the traffic sign recognition task using the \texttt{BTSR} dataset are illustrated in Figure~\ref{fig:acc-poison-gradcam-training} where \textit{weak penalization} denotes the Trojan models trained with a small $\gamma$ in Eq.~(\ref{eq:adaptive_trojan}) and  \textit{strong penalization} denotes models trained with a large $\gamma$. 

\vspace{1mm}
We make the following observations from our results in Figure~\ref{fig:acc-poison-gradcam-training}:

\vspace{2mm}
\noindent\textbf{Observation 1}. \textit{Due to the contradictory objectives between concealing the salience of trigger features (or network bias) and achieving state-of-the-art results, increasing GradCAM knowledge in the training process of a Trojaned network will degrade the classification accuracy whilst leading the network to neglect the effect of the Trojan (lower attack success rate). Achieving optimality in both objectives will lead to degrading both the attack success (ASR) and model performance (Accuracy). Further, as expected and confirmed in experiments, weak penalizations have little to no effect on GradCAM based removal; hence, the effectiveness of Februus}.

\vspace{2mm}
\noindent\textbf{Observation 2}. \textit{The average probability of predictions we obtained from the adaptively Trojaned networks---that is $\frac{1}{n}\sum_{i=1}^n p(y=c|\bx_i)$ where c is the predicted label and given as Confidence in Figure~\ref{fig:acc-poison-gradcam-training})---reduced significantly to below 20\% as we increased $\gamma$ (i.e increasing the contribution of the GradCAM loss term in~\eqref{eq:adaptive_trojan}). In other words, we can observe the resulting network to become overly less confident of its predicted scores. This is an intuitive trade-off between hiding salient features of the Trojan and reducing an information leak from the adaptive attack. Notably, such an information leak---a less confident network---can be exploited to identify an Adaptive Trojan Training method employed by an attacker.}
Interestingly, we observed similar trends on visual tasks when we attempted different adaptive training techniques such as forcing GradCAM to focus away from the Trojan region and forcing GradCAM output to be \textit{random}.

\vspace{3mm}
{\noindent\textbf{\textit{GradCAM Evasion Attack (Input Perturbations)}}. We consider an attacker attempting to fool GradCAM at \textit{inference} time. Theoretically, GradCAM can be fooled by perturbing the input with the objective of misleading the GradCAM selected input region, similar to that possible with an  \textit{adversarial example}~\cite{244036, adv_example, 43405, aleks2017deep}. Although this is out of our threat model for a Trojan attack where attackers utilize input-agnostic, realistic, natural triggers such as a tattoo, we conducted experiments to assess this threat. The results are discussed in Appendix~\ref{sec:appendix-fool-gradcam}} and illustrated in Figure~\ref{fig:attack_gradcam_overlay}. Interestingly, we observed that adding large-magnitude adversarial noise, while potentially misleading GradCAM, has the adverse effect of causing the Trojaned classifier to neglect the trigger, hence reducing the attack success rate. 

\vspace{3mm}
\noindent\textbf{\textit{GradCAM Evasion Attack (Trigger Perturbations)}}. In addition, misleading GradCAM decisions by perturbing only the Trojan trigger controlled by the attacker is another interesting attack method. Stanford researchers in a study~\cite{Chou2018SentiNetDP} have shown that localized patch perturbations only result in GradCAM focusing on the location of the trigger; thus, the possibility to mislead GradCAM by only perturbing the trigger whilst maintaining the potency of the Trojan remains an open challenge. 

\subsection{Attacks Targeting Image Restoration}
\label{sec:large_trigger}
\label{sec:target-image-restoration}

Assuming that adaptive attackers are fully aware of the Image Restoration mechanism of Februus albeit without access to manipulate the training process of the \textit{Image Restoration} module, a strong attack against the restoration is to embed a \textit{large} or \textit{multiple-piece} trigger to force an arbitrarily large removal region through \textit{Image Restoration} and to challenge the recovery during \textit{Image Restoration}.

\begin{figure}[h]
    \centering
    \includegraphics[width=\linewidth]{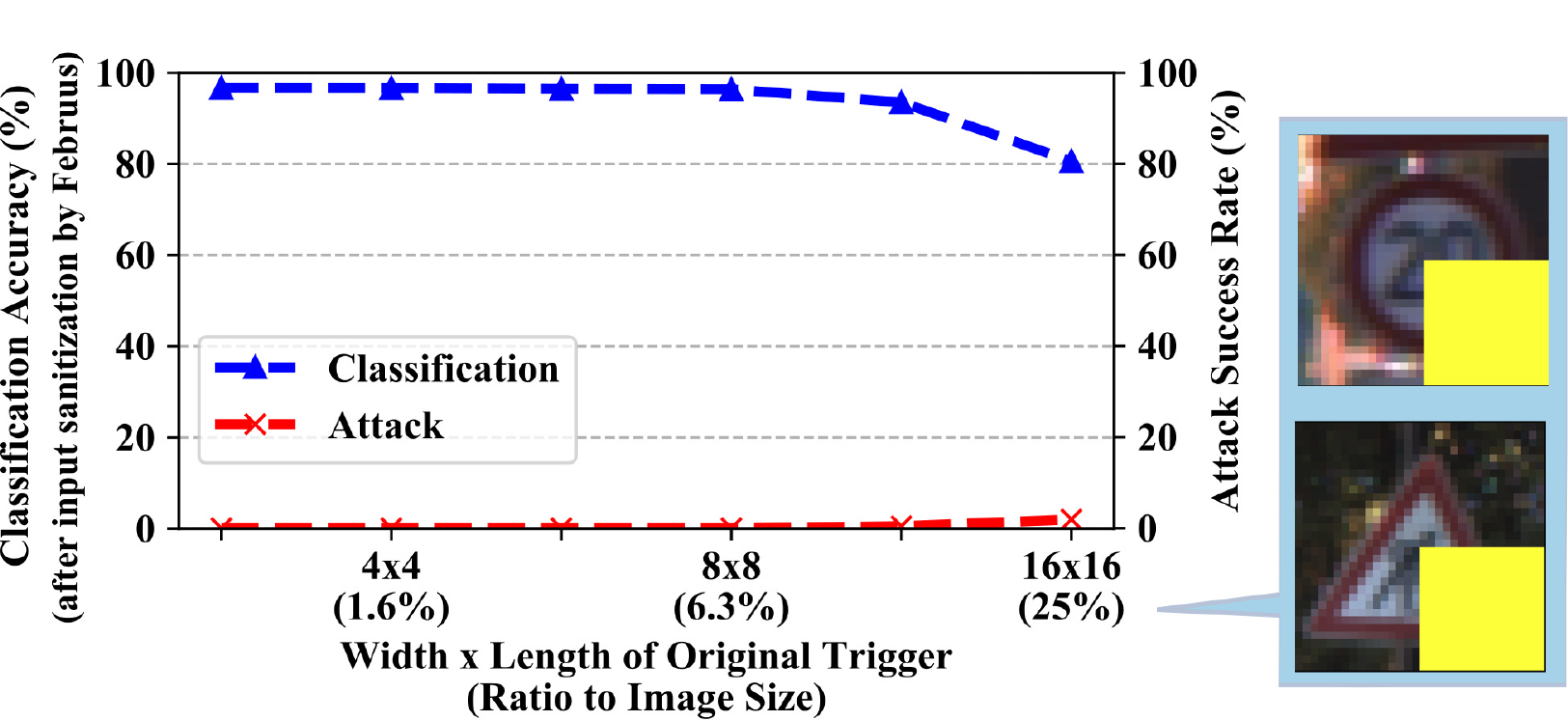}
    \caption{Februus applied to the infected GTSRB model whilst increasing the size of the Post-it trigger and illustrations of large triggers occluding 25\% of the input images. 
    } 
    \label{fig:gtsrb_trigger_size}
\end{figure}

\vspace{3mm}
\noindent\textbf{\textit{Increasing  the  trigger  size}}.  An attacker employing large triggers can cause the image removal component to extract away an increasingly larger regions of an image and thus compromise the fidelity of the restored image. The sensitivity of Februus to a larger trigger is illustrated in Figure~\ref{fig:gtsrb_trigger_size}. When the trigger covers 25\% of an image class in \texttt{GTSRB}, the attack success rate \textit{after Februus} is only 1.93\%, while it is 0\% for smaller triggers. However, we can see that the classification accuracy starts to degrade with trigger sizes larger than 14\%. As the trigger's size increases and covers up to \textit{one-fourth} of the image, the classification accuracy reduces to 80.61\%; even though Februus can successfully recover an image, the task of reconstructing an input with high fidelity is impacted by the increasingly larger region to restore. We observed similar trends in other visual tasks.

\vspace{3mm}
\noindent\textbf{Multiple-piece trigger}.\label{sec:mul_piece_1_label} An attacker can also challenge the GAN image restoration by employing a trigger with multiple pieces to force the restoration of multiple regions. With no assumptions regarding the size or the location of the Trojan during the construction of the GAN---recall that we used randomized locations and masked areas---we expect Februus to be highly generalizable to restoring multiple regions of arbitrary sizes.

\begin{figure}[h]
    \centering
        \includegraphics[width=\linewidth]{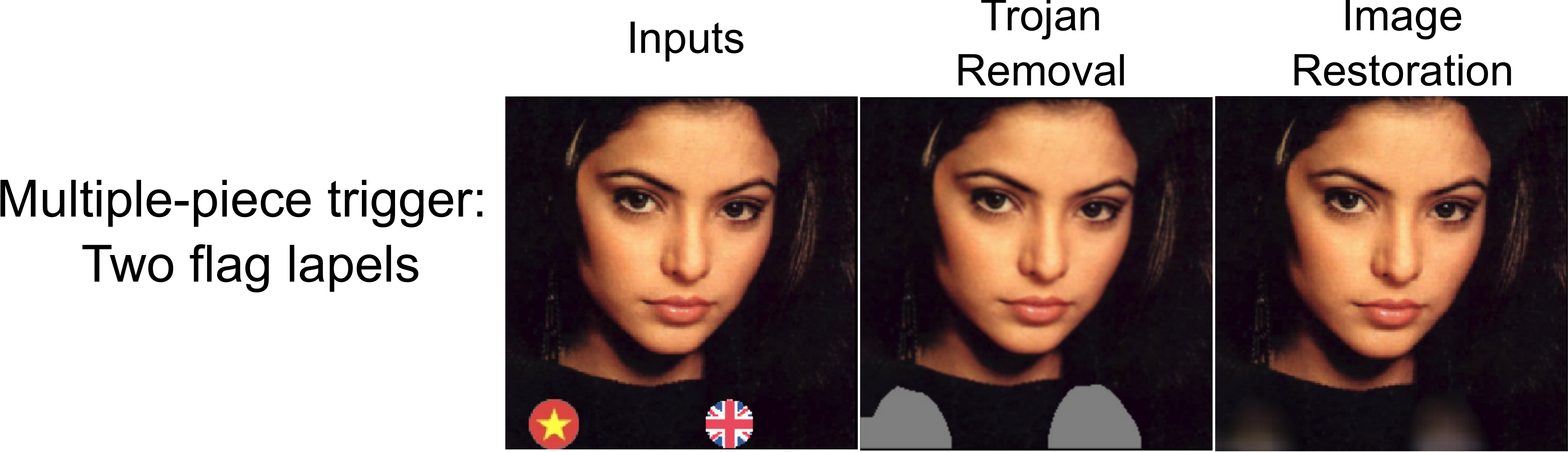}
        \caption{Multiple-piece trigger targeting a single label.}
    \label{fig:mul-piece}
\end{figure}

As shown in Table~\ref{tab:backdoor_variants}, Februus correctly identifies and eliminates all the triggers with the attack success rate reducing from 100\% to  0.32\% whilst maintaining a classification accuracy of 91.42\% for cleaned Trojaned inputs and 91.36\% for benign inputs---we illustrate a two-piece trigger example in Figure~\ref{fig:mul-piece}.

\section{Related Work and Discussion}
\label{sec:lit_comparison}

\subsection{Backdoor Attacks and Defenses}~\label{sec:related_backdoor_attack} 
\noindent\textbf{Attacks.}~Backdoor attacks have recently been recognized as a threat due to the popular trend of using pre-trained models and MLaaS. Recent works \cite{Liu2018TrojaningAO,Gu2017BadNetsIV, Bagdasaryan2018HowTB, Chen2017TargetedBA, invisiblebackdoor} have shown that attackers can embed backdoors to a ML system by poisoning the training sets with malicious samples at the training phase. While Gu et al. \cite{Gu2017BadNetsIV} assume that the attacker has full control over training where a Trojan can be of any shape or size, Chen et al.~\cite{Chen2017TargetedBA} propose an attack under a more challenging assumption where the attacker can only poison a small portion of the training set. Liu et al.~\cite{Liu2018TrojaningAO} show that they do not require the training dataset at all to Trojan a neural network and create a stealthy Trojan attack which targets dedicated neurons instead of poisoning the whole network. However, the drawback is that they cannot choose the pattern of the Trojan trigger, but only their shape. 

In addition, attempts to make a Trojan attack more stealthy, Liu et al.~\cite{liu2020reflection} presented a backdoor attack using reflections. Saha et al.~\cite{saha2019hidden} propose a novel approach to create a  backdoor by generating natural looking poisoned data with the correct ground truth labels.
On the other hand, Bagdasaryan el al.~\cite{bagdasaryan2020blind} propose a new method for injecting backdoors by poisoning the loss computation in the training code and name the method \textit{blind backdoors} since the attacker has no power to modify the training data, observe the execution of the code or the resulting models.

\vspace{3mm}
\noindent{\textbf{Defenses.}~}
 Since the attack scenarios were discovered, there has been a surge of interest in defenses against Trojan attacks \cite{AC, wang-2019-ieeesp, Chou2018SentiNetDP, Liu2018FinePruningDA, Gao2019STRIPAD, gao2019design, tabor, liu2017neural}, and some certified robustness against backdoor attacks are proposed in~\cite{zhang2020backdoor, weber2020rab, wang2020certifying}. Liu et al.~\cite{liu2017neural} proposed three methods to eliminate backdoor attacks and were evaluated on the simple MNIST dataset~\cite{mnist}. Chen et al.~\cite{AC} proposed an Activation Clustering (AC) method to detect whether the training data has been poisoned. 
This method assumes access to Trojaned inputs. Liu et al. \cite{Liu2018FinePruningDA}  developed a method named Fine-Pruning to disable backdoors by pruning DNNs and then fine-tuning the pruned network. Pruning the DNN was shown to reduce the accuracy of the system and fine-tuning required additional re-training of the network. In CCS'2019, Liu {\it et al.} proposed Artificial Brain Stimulation (ABS)~\cite{liu2019abs} to determine whether a network is Trojaned. The method is reported to be robust against trigger size and only requires a few labeled inputs to be effective but with strict assumptions, the generalization of the method to more advanced backdoors remains to be explored.

Chou et al. \cite{Chou2018SentiNetDP} and Gao et al. \cite{Gao2019STRIPAD} have proposed run-time Trojan anomaly detection methods named SentiNet and STRIP, respectively. SentiNet also utilized the GradCAM Visual Explanation tool~\cite{Selvaraju2017GradCAMVE} to understand the predictions of the DNN and detect a Trojan trigger. SentiNet also demonstrated GradCAM to be robust in identifying adversarial regions regardless of whether it is a Trojaned trigger or an adversarial patch. Gao et al. \cite{Gao2019STRIPAD} propose a backdoor anomaly detection method that can detect potential malicious inputs at run-time, which can be applied to different domains~\cite{gao2019design}. Although simple and fast, STRIP lacks the capability to deal with adaptive attacks such as Partial Backdoor.  Both of these methods focus only on Trojan detection.

In SP'2019, Wang et al. \cite{wang-2019-ieeesp} proposed Neural Cleanse, a novel idea aiming to reverse the Trojan triggers and clean a DNN and its method is further improved in~\cite{tabor}. Using reversed triggers, authors use a method of \textit{unlearning}, which requires retraining the network to patch the backdoor. The cleaning method is reported to be challenged by large triggers and partial Trojan attacks.  The idea of reversing the Trojan trigger was also proposed in DeepInspect (DI)~\cite{deepinspect} but the reported results therein, after patching the network, appear to be less favorable than Neural Cleanse. 

We provide a comparison of Februus with recent defense methods in \textbf{Appendix~\ref{sec:SOTA-compare}} and summarize our findings in Table~\ref{tab:compare_works}.

\subsection{Run-time Overhead Comparisons}
\label{sec:runtime-overhead}
Since Februus is plugged as an overhead to an existing DNN to sanitize Trojan inputs, the run-time of the Februus system should be evaluated. As shown in Table~\ref{tab:runtime}, the run-time of the entire pipeline only takes 29.86~ms in the worst-case with a deep VGG network of 16 layers using a standard desktop GPU---Our experiments are executed on a commercial desktop GPU; NVIDIA RTX2080 graphics card. 

In simpler classification tasks, the overhead is only around 6~ms or 8~ms. This result is around 800$\times$ faster than SentiNet~\cite{Chou2018SentiNetDP} which takes around 23.3s for the same task and is comparable with the fast and simple Trojan \textit{\textbf{detection only}} method in STRIP~\cite{Gao2019STRIPAD}. More importantly, the acceptable latency for autonomous driving systems from Google, Uber or Tesla is around 100ms \cite{selfdriving}. Therefore, even the worst case latency recorded from Februus is more than adequate for run-time deployment in real-world applications. In addition, even though the camera resolution could be high, the detected images are normally are captured and cropped from a long distance to make timely decisions (see Figure 11 in~\cite{realtraffic}). For example, in a real-world Traffic sign detection and recognition system~\cite{realtraffic}, the captured size for Traffic signs ranges from 13 to 250 pixels. Notably, images of these sizes were investigated in our experiments.

\begin{table}[h!]
\centering
\small
\caption{Average run-time of different classification tasks on 100 images. Even with the high-resolution images of the Face Recognition task using a complex VGG-16 network, the total run-time of the Februus system is 29.86~ms, while the simpler scene classification task only incurs a 6.32~ms overhead.}
\label{tab:runtime}
\begin{tabular}{@{}cc@{}}
\dtoprule
\textbf{Task/Dataset}                                                      & \textbf{Run-time Overhead} \\ \midrule
\begin{tabular}[c]{@{}c@{}}Scene Classification (\texttt{CIFAR10})\end{tabular}   & 6.32 ms                    \\ \midrule
\begin{tabular}[c]{@{}c@{}}German Traffic Sign Recognition (\texttt{GTSRB})\end{tabular} & 8.01 ms                    \\ \midrule
\begin{tabular}[c]{@{}c@{}}Belgium Traffic Sign Recognition (\texttt{BTSR})\end{tabular} & 6.49 ms                    \\ \midrule
\begin{tabular}[c]{@{}c@{}}Face Recognition (\texttt{VGGFace2})\end{tabular}      & 29.86 ms                   \\ \dbottomrule
\end{tabular}
\end{table}

\subsection{Limitations}
We quantitatively and qualitatively compare Februus with other state-of-the-art defense methods in \textbf{Appendix~\ref{sec:SOTA-compare}}. Februus is robust against input-agnostic Trojan attacks---our primary aim under our threat model---whilst generalizing well across complex adaptive attacks, we observed some limitations. 

Interestingly, in Section~\ref{sec:gradcam-training}, our investigations into adaptive training methods demonstrate a possibility to evade Trojan removal but we observed this to come at the cost of further information leaks or significantly degraded attack success rates.

As demonstrated in Section~\ref{sec:large_trigger}, a large trigger covering more than one-fourth of an image can cause a degradation in the classification accuracy by attacking the image restoration stage of Februus; although, Februus can successfully block the attack. 

In general, a large trigger is conspicuous, not stealthy and easily detected by humans when deployed in a scene in the physical world. For example, we illustrate in Figure~\ref{fig:gtsrb_trigger_size} the trigger required to achieve a digitization of an image with a trigger size covering 25\% of the image; hence such attacks are extremely difficult to mount. 

Further, large triggers are a challenging problem and cause a degradation in state-of-the-art Trojan defense methods. However, in comparison with 2019 IEEE S\&P Neural Cleanse method~\cite{wang-2019-ieeesp}, Februus is demonstrated to be less sensitive to these larger triggers as shown in Figure~\ref{fig:gtsrb_trigger_size} and compared to in Table~\ref{tab:compare_works} (in the Appendix~\ref{sec:SOTA-compare}). Februus can be improved by enhancing the image restoration module, for example, by using more unlabeled data to increase the fidelity of the reconstruction by the GAN or training the GAN with labeled data with the additional objective of maximizing the classification accuracy of the classifier on inpainted images to boost the performance of the GAN to maintain the classification accuracy of restored inputs. In addition, as we illustrated in Section~\ref{sec:large_trigger}, mounting such a large trigger attack in the physical world is a challenging proposition.

\section{Conclusion}
\label{sec:discussion}

The Februus has constructively turned the strength of the input-agnostic Trojan attacks into a weakness. This allows us to remove the Trojan via the bias of network decision and cleanse the Trojan effects out of malicious inputs at run-time without prior knowledge of poisoned networks and the Trojan triggers. Extensive experiments on various classification tasks have shown the robustness of our method to defend against input-agonist backdoor attacks as well as advanced variants of backdoor and adaptive attacks.  

Overall, in contrast to prior works, Februus is the first method to leverage cheaply available unlabeled data and cleaning out the Trojaned triggers from malicious inputs and patching the performance of a poisoned DNN without retraining. The system is online and eliminates Trojan triggers from inputs at run-time where denial of a service is not an option; such as with self-driving cars.

Future work should investigate the generality of the concept we first proposes and demonstrated here---input sanitization---to other domains such as speech and text. Further, whilst GradCAM and the GAN based components we employed are only one set of methods to achieve input sanitization, alternatives may prove more robust, effective or impose even a smaller run-time overhead.

\bibliographystyle{IEEEtran}  
\bibliography{Februus}

\appendices
\section{GradCAM Evasion Attacks}
\label{sec:appendix-fool-gradcam}

Besides adding GradCAM knowledge during the training process, some adaptive attackers may attempt to mislead GradCAM to propose a wrong location at the inferencing stage, and thus reduce the robustness of our defense method. Based on our extensive experiments, GradCAM is shown to be insensitive to sizes and locations of Trojan triggers (as shown in Figure~\ref{fig:trigger_locations}). 

\begin{figure}[h]
    \centering
    \includegraphics[width=0.68\linewidth]{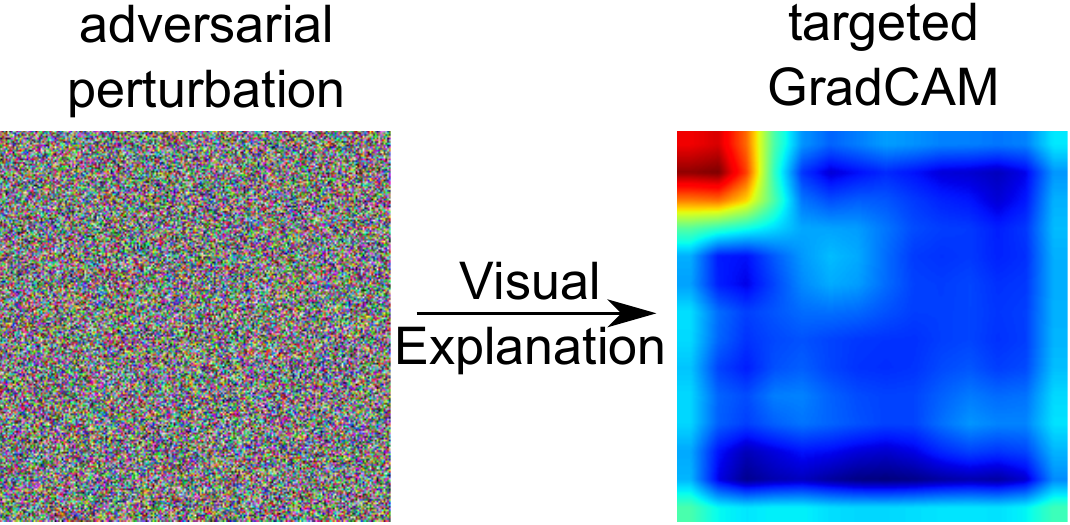}
    \caption{Adaptive Attacks on GradCAM. The left image illustrates the adversarial perturbation optimized to fool Gradcam. The right picture shows that GradCAM is fooled to detect the targeted region.}
    \label{fig:attack_gradcam}
\end{figure}

\begin{figure}[h!]
\centering
\includegraphics[width=\linewidth]{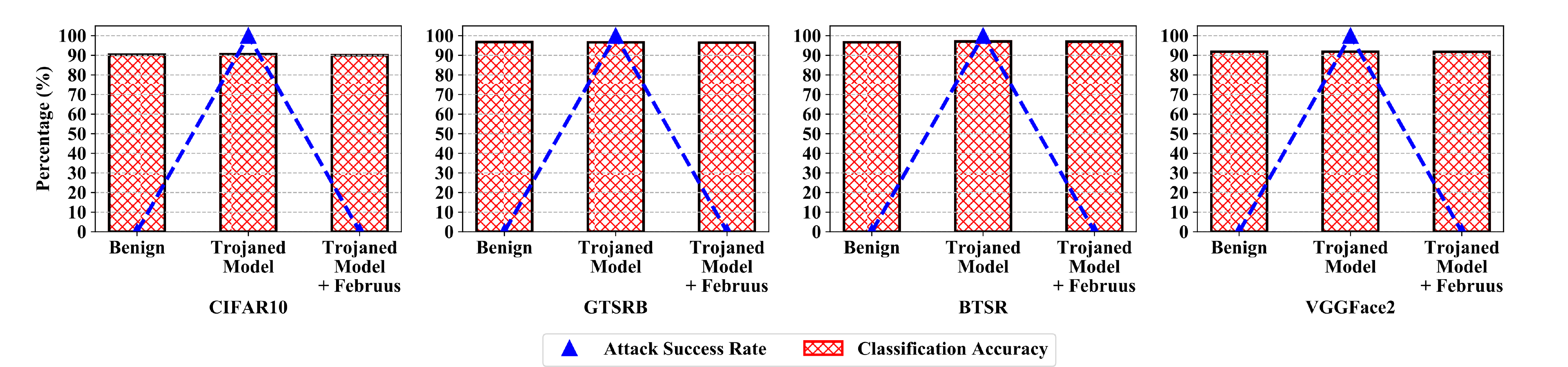}
\caption{Robustness of Februus on Different Classification Tasks. Februus is highly effective and perform consistently well against backdoor attacks. We can observe attack success rate reductions from 100\% to nearly 0\% while the classification accuracy is maintained in across the three different classification tasks. Notably, model performance after deploying Februus remains similar to that obtained from  benign inputs.} 
\label{fig:result_charts}
\end{figure}

Nevertheless, for an \textit{evasion} attack at the inferecing stage, we assume an attacker is capable of adding noise to the input scene to be digitized by a camera to fool GradCAM and misdirect to a targed region of the input. Results from this experiment are shown in Figure~\ref{fig:attack_gradcam}. Notably such an attack requires adding noise to the \textit{entire scene to be digitized} or the input image.

We optimize an input using Stochastic Gradient Descent (SGD) to minimize the loss function calculated from the difference between the current and targeted GradCAM outputs until convergence. As shown in Figure~\ref{fig:attack_gradcam}, an attacker may create a perturbation that can fool GradCAM to detect a designated region. Adaptive attackers might add this noise to the Trojaned input (with the hyper-parameter of $\epsilon$ to alter the magnitude of the noise added) to mislead GradCAM and reduce the robustness of our method (as shown in Figure~\ref{fig:attack_gradcam_overlay}). However, adding noise to the Trojaned inputs does not guarantee the ability of the Trojan to still trigger the DNN; further, this attack method is out of our threat model focusing on physically realizable Trojan triggers. 

\captionsetup[figure]{font=small}

We also recognize that a stealthy attacker may attempt to deploy perturbations within the Trojan trigger to create an adversarial trigger to attempt to fool GradCAM. However, researchers in Stanford~\cite{Chou2018SentiNetDP} showed the infeasibility of this method to fool GradCAM, unless an attacker is capable of perturbing the whole image as shown in  Figure~\ref{fig:attack_gradcam_overlay} and Figure~\ref{fig:attack_gradcam}. 

\section{Detailed Information On Datasets, Model Architectures and Training Configurations}
\label{sec:network-config}

\begin{table}[h]
\centering
\caption{Model Architecture for CIFAR-10. FC: fully-connected layer.}
\label{tab:cifar10_arch}
\begin{adjustbox}{width=\linewidth, center}
\begin{tabular}{ccccc}
\hline
Layer Type & \# of Channels & Filter Size & Stride & Activation \\ \hline
Conv       & 128            & 3           & 1      & ReLU       \\
Conv       & 128            & 3           & 1      & ReLU       \\
MaxPool    & 128            & 2           & 2      & -          \\
Conv       & 256            & 3           & 1      & ReLU       \\
Conv       & 256            & 3           & 1      & ReLU       \\
MaxPool    & 256            & 2           & 2      & -          \\
Conv       & 512            & 3           & 1      & ReLU       \\
Conv       & 512            & 3           & 1      & ReLU       \\
MaxPool    & 512            & 2           & 2      & -          \\
FC         & 1024           & -           & -      & ReLU       \\
FC         & 10             & -           & -      & Softmax    \\ \hline
\end{tabular}
\end{adjustbox}
\end{table}

\begin{table}[h]
\centering
\caption{Model Architecture for GTSRB}
\label{tab:gtsrb_arch}
\begin{adjustbox}{width=\linewidth, center}
\begin{tabular}{ccccc}
\hline
Layer Type & \# of Channels & Filter Size & Stride & Activation \\ \hline
Conv       & 128            & 3           & 1      & ReLU       \\
Conv       & 128            & 3           & 1      & ReLU       \\
MaxPool    & 128            & 2           & 2      & -          \\
Conv       & 256            & 3           & 1      & ReLU       \\
Conv       & 256            & 3           & 1      & ReLU       \\
MaxPool    & 256            & 2           & 2      & -          \\
Conv       & 512            & 3           & 1      & ReLU       \\
Conv       & 512            & 3           & 1      & ReLU       \\
MaxPool    & 512            & 2           & 2      & -          \\
Conv       & 1024           & 3           & 1      & ReLU       \\
MaxPool    & 1024           & 2           & 2      & -          \\
FC         & 1024           & -           & -      & ReLU       \\
FC         & 10             & -           & -      & Softmax    \\ \hline
\end{tabular}
\end{adjustbox}
\end{table}

\newpage
\onecolumn

\begin{table}[h]
\centering
\caption{Model Architecture for VGGFace2}
\label{tab:vggface2_arch}
\begin{adjustbox}{width=0.5\linewidth, center}
\begin{tabular}{ccccc}
\hline
Layer Type & \# of Channels & Filter Size & Stride & Activation \\ \hline
Conv       & 64             & 3           & 1      & ReLU       \\
Conv       & 64             & 3           & 1      & ReLU       \\
MaxPool    & 64             & 2           & 2      & -          \\
Conv       & 128            & 3           & 1      & ReLU       \\
Conv       & 128            & 3           & 1      & ReLU       \\
MaxPool    & 128            & 2           & 2      & -          \\
Conv       & 256            & 3           & 1      & ReLU       \\
Conv       & 256            & 3           & 1      & ReLU       \\
Conv       & 256            & 3           & 1      & ReLU       \\
MaxPool    & 256            & 2           & 2      & -          \\
Conv       & 512            & 3           & 1      & ReLU       \\
Conv       & 512            & 3           & 1      & ReLU       \\
Conv       & 512            & 3           & 1      & ReLU       \\
MaxPool    & 512            & 2           & 2      & -          \\
Conv       & 512            & 3           & 1      & ReLU       \\
Conv       & 512            & 3           & 1      & ReLU       \\
Conv       & 512            & 3           & 1      & ReLU       \\
MaxPool    & 512            & 2           & 2      & -          \\
FC         & 4096           & -           & -      & ReLU       \\
FC         & 4096           & -           & -      & ReLU       \\
FC         & 170            & -           & -      & Softmax    \\ \hline
\end{tabular}
\end{adjustbox}
\end{table}

\begin{table}[h!]
\caption{Dataset and Training Configuration}
\label{tab:training_config}
\begin{adjustbox}{width=\linewidth, center}

\begin{tabular}{|c|c|c|c|c|c|}
\hline
\textbf{Task/Dataset} & \textbf{\# of Labels} & \textbf{Input Size} & \textbf{Training Set Size} & \textbf{Testing Set Size} & \textbf{Training Configuration}                                                                                                                                                                          \\ \hline
CIFAR-10              & 10 & $32\times32\times3$                    & 50,000                     & 10,000                   & \makecell{inject ratio=0.1, epochs=100, batch=32, \\optimizer=Adam, lr=0.001}                                                                                                                                         \\ \hline
GTSRB                 & 43  & $32\times32\times3$                  & 35,288                     & 12,630                    & \makecell{inject ratio=0.1, epochs=25, batch=32, \\optimizer=Adam, lr=0.001}                                                                                                                                          \\ \hline
BTSR                 & 62 & $224\times224\times3$                   & 4,591                     & 2,534                    & \makecell{inject ratio=0.1, epochs=25, batch=32, \\optimizer=Adam, lr=0.001 }                                                                                                                                         \\ \hline
VGGFace2              & 170  & $224\times224\times3$                 & 48,498                     & 12,322                    & \begin{tabular}[c]{@{}c@{}}inject ratio=0.1, epochs=15, batch=32, \\optimizer=Adadelta, lr=0.001\\ First 10 layers are frozen during training. \\First 5 epochs are trained using clean data only.\end{tabular} \\ \hline
\end{tabular}
\end{adjustbox}
\end{table}

\noindent\textbf{GAN training algorithm}. In this section, we discuss further details of the training algorithm for Generative Adversarial Network mentioned in Section~\ref{sec:methodology}. The details are mentioned in Alg.~\ref{alg:GAN}. 

\begin{algorithm}[h!]
\caption{Training procedure for the image inpainting GAN network (with generator parameters $\btheta$).}
\label{alg:GAN}
\begin{algorithmic}[1]
\Require The gradient penalty coefficient $\lambda$, Adam optimizer hyper-parameters $\alpha$, $\beta_1$, $\beta_2$, the number of discriminator iterations per generator iteration $n_\text{critic}$, the batch size $m$, and the regularization hyperparameter of Generator loss $\gamma$.

\While{$\btheta$ has not converged}
    \For {$t=1,...,n_\text{critic}$}
        \For {$i=1,...,m$}
            \State Sample real data $\bx\sim \mathbb{P}_r$
            \State Generate a mask $\textbf{M}_c$ for $\bx$ with an arbitrary mask at randomized region and shape.
            \State Generate a masked input $G(\bx,\textbf{M}_c)$ 
            \State Get the inpainted sample $\Tilde{\bx}\sim \mathbb{P}_g$ based on the masked input $G(\bx,\textbf{M}_c)$. 
        \EndFor
        \State Update the discriminators $D$ with the joint loss gradients (Eq.~\ref{eq:loss_D}) using a batch of real data $\bx$ and inpainted data $\Tilde{\bx}$.
    \EndFor
    \State Sample a batch of real data $\bx\sim \mathbb{P}_r$
    \State Generate a mask $\textbf{M}_c$ for $\bx$ with an arbitrary mask at randomized region and shape.
    \State Generate masked data $G(\bx,\textbf{M}_c)$ 
    \State Get the inpainted samples $\Tilde{\bx}\sim \mathbb{P}_g$ based on the masked inputs $G(\bx,\textbf{M}_c)$
    \State Update the Generator $G$ with the joint loss gradients (Eq.~\ref{eq:loss_G_joint}).
\EndWhile

\end{algorithmic}
\end{algorithm}

\section{Comparison with State-of-the-art Methods}
\label{sec:SOTA-compare}
We compare ours with recently published state-of-the-art defense methods in the literature as summarized in Table~\ref{tab:compare_works}. DeepInspect~\cite{deepinspect}, Fine-pruning~\cite{Liu2018FinePruningDA}, ABS~\cite{liu2019abs} and Neural Cleanse \cite{wang-2019-ieeesp} work offline, i.e. they will perform Trojan detection in the network and patch it when it is not actively used; in contrast, Februus is online, removes and patches the inputs at run-time. 

STRIP~\cite{Gao2019STRIPAD}, akin to our approach, works in the input domain and at run-time. However, there are some differences in our method compared with theirs. The first and obvious difference is that this method only detects potential Trojans, while our method cleans the inputs. Hence, our cleaning method results should be compared with network patching results in Neural Cleanse~\cite{wang-2019-ieeesp}, or DeepInspect~\cite{deepinspect} defenses since these methods also attempt to clean the Trojaned effect whilst aiming to achieve state-of-the-art performance from the sanitized network. The second difference is that our GAN inpainting method is unsupervised, hence, we can utilize a huge amount of cheap unlabeled data to improve our defense, while other methods rely on ground-truth labeled data---both difficult and expensive to obtain. Third, our method is robust to Partial Trojan attacks and multiple triggers, two challenging attacks for our counterparts \cite{Gao2019STRIPAD,wang-2019-ieeesp,tabor}. Notably, Februus can cleanse the Trojan effects in just a single run (or pass). 

\begin{table}[h!]
\centering
\small
\begin{adjustbox}{width=\linewidth, center}
\begin{threeparttable}
\caption{Comparison between Februus and other Trojan defense methods}
\label{tab:compare_works}
\begin{tabular}{ccccccc}
\dtoprule
\textbf{Work}               & \textbf{\makecell{Costly Labeled\\Data Required}} & \textbf{Run-time} & \textbf{\makecell{DNN\\Restoration\\Capability}} & \textbf{Domain}  & \textbf{\makecell{Against Complex\\Partial Backdoor Attacks\tnote{2}}} & \textbf{\makecell{Results\\After Restoration\tnote{1}}}  \\ \midrule
\textit{SentiNet}~\cite{Chou2018SentiNetDP}             & Yes & Yes      & No                         & Input & Not Evaluated & Not Available                                    \\ \midrule
\textit{STRIP}~\cite{Gao2019STRIPAD}             & Yes & Yes      & No                         & Input & Not Capable  & Not Applicable                                    \\ \midrule
\textit{ABS}~\cite{liu2019abs}          & Yes   & No      &  No                         & Network & \makecell{Not Capable} & Not Applicable                                  \\ \midrule
\textit{DeepInspect}~\cite{deepinspect}          & No   & No      & Yes                         & Network & \makecell{Not Quantitatively\\Evaluated} & \makecell{Attack Success: 3\%, \\Classification Accuracy: 97.1\%}                                  \\ \midrule

\textit{Neural Cleanse}~\cite{wang-2019-ieeesp}   & Yes      & No       & Yes                        & Network & \makecell{Not Quantitatively\\Evaluated} & \makecell{Attack Success: 0.14\%, \\Classification Accuracy: 92.91\%, \\Cannot detect the trigger sizes \\larger than $8\times8$ }   \\ \midrule
\textit{Februus (Ours)} & No     & Yes      & Yes                        & Input & \makecell{Yes\\(in just a single run)}  & \makecell{Attack Success: 0.00\%,\\ Classification Accuracy: 96.64\%, \\Can block the Trojan effect \\with large trigger size of\\ $16\times16$ (cover 25\% of the picture).} 
        \\ \dbottomrule
\end{tabular}
\begin{tablenotes}
\item[1] The comparison is on the GTSRB dataset shared by all methods in respective experimental evaluations. Notably, the classification accuracy of the methods we compare with are after the model is re-trained using clean labeled data. 
\item[2] The methods that discuss potential defenses require adapting their defense mechanisms and knowledge of trojaning implementations; notably, such information may be difficult to gain in practice. 
\end{tablenotes}
\end{threeparttable}
\end{adjustbox}
\end{table}

\end{document}